\documentclass[prd,twocolumn,superscriptaddress,showpacs,amssymb,amsmath,
amsfonts,aps,altaffilletter]{revtex4}

\usepackage{color}
\usepackage{times}
\usepackage{graphicx}
\usepackage{fancyhdr}
\usepackage{float}
\usepackage{ulem}
\makeatletter
\makeatletter
\def\@fnsymbol#1{\ifcase#1\or * \or  $+$ \or  \$ \or \#  \or \dag \or \ddag \or
$\mathsection$ \or $ \mathparagraph$ \or $\|$  \or \textordfeminine \or \textbul
let   
\or ** \or $++$ \or  \$\$ \or \#\#  \or \dag\dag \or \ddag\ddag \or
$\mathsection\mathsection$ \or $ \mathparagraph\mathparagraph$ \or $\|\|$  \or 
\textordfeminine\textordfeminine \or \textbullet \textbullet \or *** \or $+++$ 
\or  \$\$\$ \or \#\#  \or \dag\dag \or \ddag\ddag \or
$\mathsection \mathsection\mathsection$ \or $ \mathparagraph 
\mathparagraph\mathparagraph$ \or $\|\|\|$  \or 
\textordfeminine\textordfeminine\textordfeminine \or 
\textbullet\textbullet\textbullet \or \else \@ctrerr\fi}
\makeatother

\def\thercsid{\relax}
\def\rcsid#1{\def\next##1#1{\def\thercsid{##1}}\next}
\rcsid$Id: paper.tex,v 1.32 2008/06/27 11:04:36 gareth Exp $

\renewcommand{\today}{\number\day\space\ifcase\month\or
  January\or February\or March\or April\or May\or June\or
  July\or August\or September\or October\or November\or December\fi
  \space\number\year}

\begin{document}

\title{Search of S3 LIGO data for gravitational wave signals from spinning black hole and
neutron star binary inspirals
%\\\large
%{\color{red}{\large LIGO-P070102-06-Z}\\\large
%Circulation restricted to LIGO-I members}
}

% S3 author list LIGO doc T050096-04-Z
%\documentclass[prd,preprintnumbers,superscriptaddress,showpacs,amssymb,amsmath,amsfonts,aps,altaffilletter]{revtex4}
%\def\rcsfile#1{\def\next##1#1{\def\thercsfile{##1}}\next}
%\def\rcsrevision#1{\def\next##1#1{\def\thercsrevision{##1}}\next}
%\def\rcsname#1{\def\next##1#1{\def\thercsname{##1}}\next}
%\def\rcsdate#1{\def\next##1#1{\def\thercsdate{##1}}\next}
%\makeatletter
%\def\@fnsymbol#1{\ifcase#1\or * \or  $+$ \or  \$ \or \#  \or \dag \or \ddag \or
%$\mathsection$ \or $ \mathparagraph$ \or $\|$  \or
%\textordfeminine \or \textbullet   \or ** \or $++$ \or  \$\$ \or
%\#\#  \or \dag\dag \or \ddag\ddag \or $\mathsection\mathsection$
%\or $ \mathparagraph\mathparagraph$ \or $\|\|$  \or
%\textordfeminine\textordfeminine \or \textbullet \textbullet \or
%*** \or $+++$ \or  \$\$\$ \or \#\#  \or \dag\dag \or \ddag\ddag
%\or $\mathsection \mathsection\mathsection$ \or $ \mathparagraph
%\mathparagraph\mathparagraph$ \or $\|\|\|$  \or
%\textordfeminine\textordfeminine\textordfeminine \or
%\textbullet\textbullet\textbullet \or \else \@ctrerr\fi}
%\makeatother
%\begin{document}
%\rcsfile$RCSfile: authorlist_s3.tex,v $
%\rcsrevision$Revision: 1.2 $
%\rcsname$Name:  $
%\rcsdate$Date: 2007/09/24 10:06:07 $
%\preprint{\thercsfile \thercsrevision \thercsdate \thercsname\ T060106-10}
%\title{S5 LIGO Scientific Collaboration Author List\\{\normalfont\small Generated from Author Database Revision: 1.16  Date: 2007/08/02 15:01:15  Name: S5-V03 }}
\affiliation{Albert-Einstein-Institut, Max-Planck-Institut f\"ur Gravitationsphysik, D-14476 Golm, Germany}
\affiliation{Albert-Einstein-Institut, Max-Planck-Institut f\"ur Gravitationsphysik, D-30167 Hannover, Germany}
\affiliation{Andrews University, Berrien Springs, MI 49104 USA}
\affiliation{Australian National University, Canberra, 0200, Australia}
\affiliation{California Institute of Technology, Pasadena, CA  91125, USA}
\affiliation{Caltech-CaRT, Pasadena, CA  91125, USA}
\affiliation{Cardiff University, Cardiff, CF24 3AA, United Kingdom}
\affiliation{Carleton College, Northfield, MN  55057, USA}
\affiliation{Charles Sturt University, Wagga Wagga, NSW 2678, Australia}
\affiliation{Columbia University, New York, NY  10027, USA}
\affiliation{Embry-Riddle Aeronautical University, Prescott, AZ   86301 USA}
\affiliation{Hobart and William Smith Colleges, Geneva, NY  14456, USA}
\affiliation{Inter-University Centre for Astronomy  and Astrophysics, Pune - 411007, India}
\affiliation{LIGO - California Institute of Technology, Pasadena, CA  91125, USA}
\affiliation{LIGO Hanford Observatory, Richland, WA  99352, USA}
\affiliation{LIGO Livingston Observatory, Livingston, LA  70754, USA}
\affiliation{LIGO - Massachusetts Institute of Technology, Cambridge, MA 02139, USA}
\affiliation{Louisiana State University, Baton Rouge, LA  70803, USA}
\affiliation{Louisiana Tech University, Ruston, LA  71272, USA}
\affiliation{Loyola University, New Orleans, LA 70118, USA}
\affiliation{Moscow State University, Moscow, 119992, Russia}
\affiliation{NASA/Goddard Space Flight Center, Greenbelt, MD  20771, USA}
\affiliation{National Astronomical Observatory of Japan, Tokyo  181-8588, Japan}
\affiliation{Northwestern University, Evanston, IL  60208, USA}
\affiliation{Rochester Institute of Technology, Rochester, NY 14623, USA}
\affiliation{Rutherford Appleton Laboratory, Chilton, Didcot, Oxon OX11 0QX United Kingdom}
\affiliation{San Jose State University, San Jose, CA 95192, USA}
\affiliation{Southeastern Louisiana University, Hammond, LA  70402, USA}
\affiliation{Southern University and A\&M College, Baton Rouge, LA  70813, USA}
\affiliation{Stanford University, Stanford, CA  94305, USA}
\affiliation{Syracuse University, Syracuse, NY  13244, USA}
\affiliation{The Pennsylvania State University, University Park, PA  16802, USA}
\affiliation{The University of Texas at Brownsville and Texas Southmost College, Brownsville, TX  78520, USA}
\affiliation{Trinity University, San Antonio, TX  78212, USA}
\affiliation{Universitat de les Illes Balears, E-07122 Palma de Mallorca, Spain}
\affiliation{Universit\"at Hannover, D-30167 Hannover, Germany}
\affiliation{University of Adelaide, Adelaide, SA 5005, Australia}
\affiliation{University of Birmingham, Birmingham, B15 2TT, United Kingdom}
\affiliation{University of Florida, Gainesville, FL  32611, USA}
\affiliation{University of Glasgow, Glasgow, G12 8QQ, United Kingdom}
\affiliation{University of Maryland, College Park, MD 20742 USA}
\affiliation{University of Michigan, Ann Arbor, MI  48109, USA}
\affiliation{University of Oregon, Eugene, OR  97403, USA}
\affiliation{University of Rochester, Rochester, NY  14627, USA}
\affiliation{University of Salerno, 84084 Fisciano (Salerno), Italy}
\affiliation{University of Sannio at Benevento, I-82100 Benevento, Italy}
\affiliation{University of Southampton, Southampton, SO17 1BJ, United Kingdom}
\affiliation{University of Strathclyde, Glasgow, G1 1XQ, United Kingdom}
\affiliation{University of Washington, Seattle, WA, 98195}
\affiliation{University of Western Australia, Crawley, WA 6009, Australia}
\affiliation{University of Wisconsin-Milwaukee, Milwaukee, WI  53201, USA}
\affiliation{Washington State University, Pullman, WA 99164, USA}
\author{B.~Abbott}\affiliation{LIGO - California Institute of Technology, Pasadena, CA  91125, USA}
\author{R.~Abbott}\affiliation{LIGO - California Institute of Technology, Pasadena, CA  91125, USA}
\author{R.~Adhikari}\affiliation{LIGO - California Institute of Technology, Pasadena, CA  91125, USA}
\author{J.~Agresti}\affiliation{LIGO - California Institute of Technology, Pasadena, CA  91125, USA}
\author{P.~Ajith}\affiliation{Albert-Einstein-Institut, Max-Planck-Institut f\"ur Gravitationsphysik, D-30167 Hannover, Germany}
\author{B.~Allen}\affiliation{Albert-Einstein-Institut, Max-Planck-Institut f\"ur Gravitationsphysik, D-30167 Hannover, Germany}\affiliation{University of Wisconsin-Milwaukee, Milwaukee, WI  53201, USA}
\author{R.~Amin}\affiliation{Louisiana State University, Baton Rouge, LA  70803, USA}
\author{S.~B.~Anderson}\affiliation{LIGO - California Institute of Technology, Pasadena, CA  91125, USA}
\author{W.~G.~Anderson}\affiliation{University of Wisconsin-Milwaukee, Milwaukee, WI  53201, USA}
\author{M.~Arain}\affiliation{University of Florida, Gainesville, FL  32611, USA}
% ...
\collaboration{The LIGO Scientific Collaboration, http://www.ligo.org}
\noaffiliation

%\date{\today}
%\begin{abstract}
%This author list contains all authors and institution names for LSC papers. For
%minor corrections (e.g\@. spelling, initials), contact Norna Robertson, Chair
%of the LSC Election and Membership committee (robertson\_n@ligo.caltech.edu)
%For all other queries please contact your PI or relevant authorship list
%organiser for your institution in the first instance.
%\end{abstract}
%\maketitle
%\acknowledgments
%\begin{center}\itshape
%The following paragraph should serve as a starting point for
%acknowledging the support LSC has received.
%\end{center}
%The authors gratefully acknowledge the support of the United States
%National Science Foundation for the construction and operation of
%the LIGO Laboratory and the Particle Physics and Astronomy Research
%Council of the United Kingdom, the Max-Planck-Society and the State
%of Niedersachsen/Germany for support of the construction and
%operation of the GEO600 detector. The authors also gratefully
%acknowledge the support of the research by these agencies and by the
%Australian Research Council, the Natural Sciences and Engineering
%Research Council of Canada, the Council of Scientific and Industrial
%Research of India, the Department of Science and Technology of
%India, the Spanish Ministerio de Educacion y Ciencia, The National
%Aeronautics and Space Administration, the John Simon Guggenheim
%Foundation, the Alexander von Humboldt Foundation, the Leverhulme
%Trust, the David and Lucile Packard Foundation, the Research
%Corporation, and the Alfred P. Sloan Foundation.
%\end{document}

\date[\relax]{ RCS \thercsid; compiled \today }
\pacs{95.85.Sz, 04.80.Nn, 07.05.Kf, 97.80.--d}

\begin{abstract}\quad
We report on the methods and results of the first dedicated search for
gravitational waves emitted during
the inspiral of compact binaries with spinning component bodies.
We analyze
$788$ hours of data collected during the third science run (S3) of the LIGO
detectors.
We searched for binary systems using a detection template family designed
specially to capture the effects of the spin-induced precession of the
orbital plane.
We present details of the techniques developed to enable this search
for spin-modulated gravitational waves, highlighting the differences
between this and other recent searches for binaries with non-spinning
components.
The template bank we employed was found to yield high matches with our
spin-modulated target waveform for binaries with masses in the asymmetric
range
$1.0~M_{\odot} < m_{1} < 3.0~M_{\odot}$ and
$12.0~M_{\odot} < m_{2} < 20.0~M_{\odot}$
which is where we would expect the spin of the binary's components to
have significant effect.
We find that our search of S3 LIGO data had good sensitivity to
binaries in the Milky Way and to a small fraction of binaries in
M31 and M33 with masses
in the range $1.0~M_{\odot} < m_{1}, m_{2} < 20.0~M_{\odot}$.
No gravitational wave signals were identified during this search.
Assuming a binary population with
spinning components and Gaussian distribution of
masses representing a prototypical neutron star - black hole system with
$m_1 \simeq 1.35 M_{\odot}$ and $m_2 \simeq 5 M_{\odot}$,
we calculate the $90\%-$confidence upper limit on the rate of coalescence of
these systems to be
$15.9 \, \mathrm{yr}^{-1} \mathrm{L}_{10}^{-1}$,
where $\mathrm{L}_{10}$ is $10^{10}$ times the blue light luminosity of the Sun.
\end{abstract}
\maketitle

\rcsid$Id: intro.tex,v 1.47 2008/06/25 14:55:27 gareth Exp $
\section{Introduction}
\label{sub:intro}

Currently, there is a world-wide network of 
%first generation
kilometer scale
interferometric gravitational wave detectors that are either 
at or approaching their respective design sensitivities. 
The network includes the US Laser Interferometer 
Gravitational-wave Observatory 
(LIGO)~\cite{LIGOS1instpaper, Barish:1999}, 
% consistent with S3S4joint, S2BNS 
the British-German GEO600~\cite{Luck:1997hv} and 
the French-Italian Virgo~\cite{0264-9381-23-19-S01}.
%and the Japanese TAMA300~\cite{0264-9381-22-18-S02}.  
The radiation emitted during the inspiral stage of a stellar mass compact binary system 
is thought to be a likely candidate for the first direct detection of
gravitational waves using these interferometers~\cite{thorne.k:1987,grishchuk-2001-171}. 
The initial interferometers will be able to search for binary neutron
star systems as far as the Virgo Cluster, and higher mass binaries which
include black holes as far as the Coma supercluster. 
The range of merger rates consistent with present astrophysical understanding 
is summarized in Ref.~\cite{LIGOS3S4all}. 
When binary formation in star clusters is taken into account with relatively 
optimistic assumptions, detection rates could be as high as a few events per 
year for initial LIGO \cite{PortegiesZwart:2000,2006ApJForS3S4Joint,imbhlisa-2006}. 
Merger rates derived for binary populations in galactic fields consistent 
with observational constraints from the known galactic neutron star - neutron star systems
are highly uncertain but are likely to lie in the ranges (at $95\%$ confidence) 
$0.1-15 \times 10^{-6} \rm{yr}^{-1} \, \rm{L}_{10}^{-1}$ and
$0.15-10 \times 10^{-6} \rm{yr}^{-1} \, \rm{L}_{10}^{-1}$ 
for black hole - black hole and neutron star - black hole binaries respectively
\cite{OShaughnessy:2005,OShaughnessy:2006b}.
%
%The initial interferometers will be able to search for such systems well
%beyond the Virgo supercluster with an expected detectable
%rate of one inspiral event every few years~\cite{grishchuk-2001-171}.  
The LIGO Scientific Collaboration (LSC) has searched for 
compact binaries with non-spinning stellar mass components in data collected during
the first, second, third and fourth science runs (henceforth S1, S2, S3
and S4, respectively)~\cite{LIGOS2bbh,LIGOS3S4all}, by employing 
optimal matched filtering techniques~\cite{helmstrom-1968} wherein 
detector data is cross-correlated with a bank of ``templates'' 
which represent the best current knowledge of the emitted waveforms.

\begin{figure}[hbtp]
\begin{center}
\includegraphics[width=0.5\textwidth]{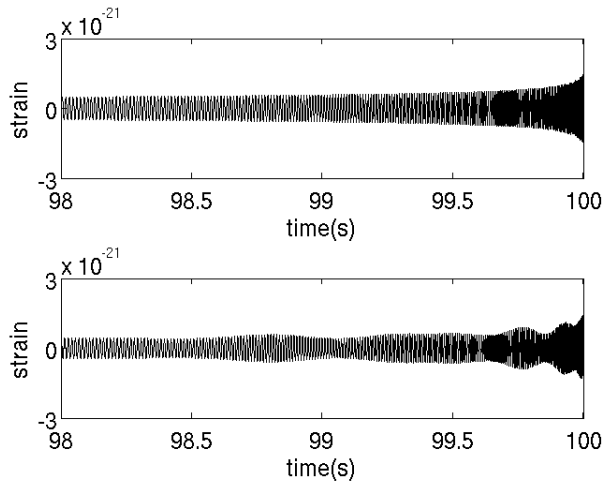}
\end{center}
\caption{The gravitational waveforms predicted from the late 
inspiral phase of two different neutron star - black hole systems, 
one consisting of non-spinning bodies (upper plot) and the other consisting of 
maximally spinning bodies (lower plot).
Both systems are identical apart from the spin of their component bodies.
Spin-induced precession of the binary's orbital plane causes modulation of
the gravitational wave signal and can be clearly seen 
in the lower plot.}
\label{waveforms}
\end{figure}

Studies of compact binaries with spinning 
components~\cite{Apostolatos:1994, kidder:821, Apostolatos:1995,Apostolatos:1996rg,
BuonannoChenVallisneri:2003b,GrandclementKalogeraVecchio:2003}
have revealed that general-relativistic dynamical coupling between the spin and orbital
angular momenta 
(so long as they are not perfectly aligned or anti-aligned)
will lead to precession of the binary's
orbital plane 
which in turn causes a modulation of the observed gravitational waves' amplitude and phase. 
The binary's orbital angular momentum $\mathbf{L}$, and therefore its orbital plane, 
and the spin angular momenta of the binary's components $\mathbf{S}_{1}, \mathbf{S}_{2}$ 
will precess about its near constant total angular momentum 
$\mathbf{J} = \mathbf{L} + \mathbf{S}_{1} + \mathbf{S}_{2}$.
The gravitational waves observed from a binary depend upon the orientation 
of the binary relative to the detector and are strongest along the direction of
its orbital angular momentum. 
The amplitude and phase of the gravitational waves emitted by the binary 
that will be observed at any particular (fixed) location will therefore
be modulated by the precession of the binary's orbital plane.
This precession of the orbital plane is nicely illustrated in Ref.~\cite{Apostolatos:1994} 
(see Fig.~2 and the appendix in particular).
Figure~\ref{waveforms} compares the gravitational waveforms we would expect to observe from 
two different binary systems, one consisting of non-spinning bodies and the other consisting 
of spinning bodies. 
The precession of the binary's orbital plane, which is related to the 
Lense--Thirring effect on gyroscopes in curved spacetimes \cite{BarkerOConnell1975}, 
should not be confused with the
in-plane precession of a binary's periastron which occurs in both spinning and non-spinning
systems. 
In this search, we consider signals with frequencies
of $70$ Hz and above, corresponding to orbital frequencies of $\ge$ 35 Hz.
We are thus sensitive only to the final stages of the binary inspiral.
By this point, the binary orbit has been circularized
due to the emission of gravitational waves (see Fig.~5 of Ref.~\cite{Belczynski:2002});
so the precession of periastron degenerates into a secular term in the evolution
of the phase.

The statistical
distribution of the spins of black holes in inspiraling binaries
is not well known~\cite{Kalogera:2000,2005ApJ...632.1035O} and until recently the efforts have focused upon
developing techniques for the detection of binary systems with
non-spinning components (for recent reviews see 
Refs.~\cite{BBCCS:2006,findchirppaper} and references therein).
The presence of amplitude and phase modulations in the
observed waveforms will reduce our detection efficiency
when using matched filter templates which do not include spin 
effects~\cite{Apostolatos:1995,Apostolatos:1996rg,BuonannoChenVallisneri:2003b,GrandclementKalogeraVecchio:2003}.
These effects are small for low mass binaries or binaries with roughly equal component masses,
but can be significant for high mass or asymmetric systems such as 
neutron star - black hole binaries.

This paper reports the methods and results of a search
for gravitational waves emitted during the inspiral 
of binaries consisting of spinning compact objects.
This search uses a detection template family designed to capture the spin-induced modulations of
the gravitational waveform which could have resulted in them being missed by other searches
targeted at non-spinning systems. 
This is the first time gravitational wave data has been searched 
for inspiral signals from binary systems with spinning component bodies.

LIGO consists of three
detectors located at two sites across the US. The LIGO Hanford Observatory
(LHO) in Washington state consists of two co-located interferometers of
arm length 4km and 2km and are known as H1 and H2 respectively.
The LIGO Livingston Observatory (LLO) in Louisiana consists of a single 4km
interferometer known as L1.
All three detectors were operated throughout S3 which spanned 70 days (1680 hours)
between October 31, 2003 and January 9, 2004.
The gravitational waves emitted by stellar mass compact binaries are expected
to be at frequencies detectable by LIGO during the final few
seconds of the inspiral as well as the merger and ringdown
stages of their evolution. 
We analyze S3 LIGO data using a detection template
family~\cite{BuonannoChenVallisneri:2003b} which efficiently 
captures the amplitude and phase modulations of the signal. 

In Sec.~\ref{evol} we discuss the evolution of spinning binary systems.
In Sec.~\ref{sub:target} we describe the waveforms
that are used to model the emission of the target sources we are seeking to detect.
These target waveforms include modulations to their amplitude and phase
in order to simulate the effects of spin-induced precession of the source. 
In Sec.~\ref{sub:dtf} we describe the detection 
template family that we use to search for these target waveforms and 
in Sec.~\ref{sub:templatebank} we describe the design and testing of the template bank
used.
In Sec.~\ref{sub:pipeline} we describe the S3 data set and summarize the 
data analysis pipeline.
In Sec.~\ref{sub:vetoes} we describe various vetoes which were identified
as beneficial to this search.
In Sec.~\ref{sub:results} we detail results from this search.
In the absence of a detection we will calculate
an upper limit on the rate of coalescences using 
the measured efficiency of our search and an estimated population 
model of the distribution of binary systems in
the Universe.
%and physical models of gravitational waves taken
%from current literature.
In Sec.~\ref{sub:upperlimit} we perform an upper limit calculation based
upon the loudest event candidate found in our search.
Finally, in Sec.~\ref{sub:conclusions} we draw conclusions. 
%* do we need a discussion of S3 instrument sensitivity in this paper? has it appeared elsewhere? Gaby
Throughout we shall assume $G = c = 1$.

\section{Evolution of spinning binary systems}
\label{evol}
We briefly review the current literature regarding the formation and evolution of spinning
binary systems.
The literature available focuses mainly on neutron star - black hole (NS-BH) binaries 
(rather than BH-BH binaries). 
Later we shall show that the template bank used in this search is most sensitive 
to binaries with unequal masses such as NS-BH binaries. 
It is likely that the formation of BH-BH and NS-BH (and indeed NS-NS) 
systems are qualitatively similar and that the discussion here will be relevant to all cases.

A typical NS-BH evolution would involve two main sequence stars in binary orbit. 
As it evolves away from the main sequence, the more massive star 
would expand until it fills its Roche lobe before
transferring mass to its companion. 
The more massive body would eventually undergo core-collapse 
to form a BH, and the system as a whole would become a high-mass X-ray binary. 
As the second body expands and evolves it would eventually fill its own Roche lobe and the binary
would then go through a common-envelope phase. This common-envelope phase, characterized by
unstable mass transfer, would be highly dissipative and would probably lead to both contraction
and circularization of the binary's orbit. 
Accretion of mass can allow the BH to spin-up. 
It has been argued that the common-envelope phase,
and associated orbital contraction, is essential in the formation of a binary which will 
coalesce within the Hubble time~\cite{Kalogera:2000}. 
Finally the secondary body would undergo core-collapse to form a NS (or if massive enough, a BH).
Prior to the supernova associated with the core-collapse of the secondary body 
we would expect the spin of the BH to be 
aligned with the binary's orbital angular momentum \cite{Kalogera:2000}.
However, the ``kick'' associated with the supernova of the secondary body could cause
the orbital angular momentum of the post-supernova binary to become tilted with respect to 
the orbital
angular momentum of the pre-supernova binary. Since the BH would have a small cross-section
with respect to the supernova kick we expect any change to the direction of its spin
angular momentum to be negligible and that the BH spin would be misaligned with respect to the
post-supernova orbital angular momentum~\cite{grandclement:102002}.
The misalignment between the spin and orbital angular momentum is expected to be preserved
until the system becomes detectable to ground based interferometers. 

The magnitude of a compact object's spin is dependent upon both its 
spin at formation (i.e., birth-spin) and the spin it attains through 
subsequent accretion episodes. 
The dimensionless spin parameter $\chi$ is given by $J/M^2$ where $J$ is the total angular
momentum of the compact object and $M$ is its mass. 
For a maximally spinning compact object we would have
$\chi=1$ and for a non-spinning object $\chi=0$. 
Although the estimated birth-spins of NS and BHs are small, simulations have
shown that accretion during a common-envelope phase can allow objects to
achieve considerable or even near maximal spins \cite{2005ApJ...632.1035O}.
Due to uncertainties in both the estimation of birth-spins and modelling
of accretion induced spin-up predictions of binary's spin population are fairly
uncertain.
The upper bound on a BH's spin is expected to be $\chi \sim 0.998$. 
Torque caused by radiation emitted from the accretion disk getting swallowed
by the BH counteracts the increase of spin caused as the BH accretes mass \cite{Thorne:1974ve}.
The upper bound of a NS's spin is estimated by calculating the spin which would cause it to
break up using variety of models for its equation of state. The upper limit is estimated to
be $\chi \sim 0.7$ \cite{Cook:1993qr}.

Techniques to measure the spin of accreting black holes using electromagnetic observations of
their accretion disk are described in Ref.~\cite{remillard-2006-44}.
Using these techniques the spins of a black holes in a handful of X-ray binaries
have been measured, in the case of GRS 1915+105 the black hole's spin was found to
be $\chi > 0.98$ \cite{mcclintock-2006-652}. 
Recent observations of spin precession through measurement of pulse shapes from binary 
radio pulsars demonstrate misalignment between the orbital and spin angular momenta of
these systems, see for example Ref.~\cite{Clifton-2008}.

For optimal detection of gravitational waves using matched-filter techniques we must 
construct templates that represent our best predictions of the signal. 
These templates must model the spin-induced modulations to the waveform's amplitude 
and phase as accurately as possible while still resulting in a computationally 
manageable number of templates covering the detectable parameter space. 
It has been shown previously~\cite{Apostolatos:1995,Apostolatos:1996rg,BuonannoChenVallisneri:2003b,GrandclementKalogeraVecchio:2003}
%\cite{Apostolatos:1994, Apostolatos:1995, Kalogera:2000} 
that if spin effects are neglected
when constructing our templates that our detection efficiency will decrease and some
spinning binary systems will be missed.
Spin effects are more pronounced when the system's spin angular momentum is larger than its orbital
angular momentum. 
%The orbital angular momentum of a binary system is approximately proportional to its mass ratio,
%meaning that spin will have more effect on systems with unequal masses
%(i.e., asymmetric systems such as NS-BH binaries, see Fig.~3 of Ref.~\cite{kidder:821}). 
The Newtonian expression for the magnitude of the orbital angular momentum 
of a binary system is $|L_{N}| = \eta M^{5/3} \omega^{-1/3}$ 
where $M = m_{1} + m_{2}$ is the total mass of the system, 
$\eta = m_{1} m_{2} / M^{2}$ is its symmetric mass ratio and
$\omega$ is the instantaneous orbital frequency of the system.
For given values of $M$ and $\omega$, the orbital angular momentum will be 
largest for binary systems with equal masses, $m_{1} = m_{2}$.
For systems with unequal masses such as NS-BH binaries, the orbital angular
momentum will be smaller and the spin angular momentum will play a more significant 
role in the system's evolution. 
It will therefore be more susceptible to the
effects of spin than equal mass systems 
(see Fig.~3 of Ref.~\cite{kidder:821}).
For schemes that fail to take into account spin effects, detection efficiency will be worse for
binaries with
i) unequal mass components,
ii) components with large spin magnitude and 
iii) when there is significant misalignment between the spins and the orbital angular momentum.

%i) the mass ratio of the binary decreases (i.e., the system becomes more asymmetric), 
%ii) the spin magnitude increases or 
%iii) the misalignment between the spin and orbital angular momentum increases.
 % Gareth and Sathya
\rcsid$Id: target_waveforms.tex,v 1.12 2007/12/07 17:57:51 gareth Exp $
\section{Target Waveforms}
\label{sub:target}

%\assign{Michele}

In this section we describe the fiducial \emph{target waveforms} used to 
represent the gravitational wave signals expected from binary systems of 
spinning compact objects. 
We adopt the post-Newtonian (PN) equations 
given out in Ref.~\cite{BuonannoChenVallisneri:2003b} and based upon 
% Ref.~\cite{kidder:821}
Refs.~\cite{kidder:821, Blanchet:1995ez, Blanchet:1996pi, PhysRevD.54.1417, Blanchet:2001ax, PhysRevD.47.R4183, Damour:2000zb, Apostolatos:1994}
(see Ref.~\cite{BuonannoChenVallisneri:2003b} for a complete list of references to
all original derivations), which model the inspiral of 
the binary in the adiabatic limit. 
In this limit the binary's components follow a sequence of 
shrinking instantaneously-circular orbits in 
a precessing orbital plane.  

The instantaneous orbital frequency $\omega$ evolves according to Eq.\ (1) 
of Ref.~\cite{BuonannoChenVallisneri:2003b}, which has the structure
\begin{equation}
\label{eq:targetomega}
\frac{\dot{\omega}(t)}{\omega(t)^2} = F_{\dot{\omega}} \bigl(
\omega(t),\hat{\mathbf{L}}_N(t) \cdot \hat{\mathbf{S}}_{1,2}(t),\hat{\mathbf{S}}_{1}(t) \cdot \hat{\mathbf{S}}_{2}(t);
M,\eta,\chi_{1,2}\bigr),
\end{equation}
with the total mass of the system $M$, the symmetric mass ratio $\eta$, 
the magnitudes of the binary's dimensionless spin parameters $\chi_{1,2}$, 
the direction of the Newtonian angular momentum  $\hat{\mathbf{L}}_N(t)$ ($\propto \mathbf{r} \times \mathbf{v}$, 
perpendicular to the bodies velocity and the vector joining them), 
and the directions of the two spins $\hat{\mathbf{S}}_{1,2}(t)$. 
Orbital PN effects are included up to 3.5PN order, while spin effects are 
included up to 2PN order.

The two spins and the orbital angular momentum evolve according to standard 
general-relativistic precession equations, which are truncated consistently 
at the relevant PN order, and which have the structure
\begin{eqnarray}
\label{eq:targetall}
\dot{\hat{\mathbf{S}}}_1 &=& \mathbf{F}_{\dot{\hat{\mathbf{S}}}_1}\!\bigl( \omega, \hat{\mathbf{L}}_N, \hat{\mathbf{S}}_2 ; M, \eta, \chi_2 \bigr) \times \hat{\mathbf{S}}_1, \nonumber \\
\dot{\hat{\mathbf{S}}}_2 &=& \mathbf{F}_{\dot{\hat{\mathbf{S}}}_2}\!\bigl( \omega, \hat{\mathbf{L}}_N, \hat{\mathbf{S}}_1 ; M, \eta, \chi_1 \bigr) \times \hat{\mathbf{S}}_2, \\
\dot{\hat{\mathbf{L}}}_N &=& \mathbf{F}_{\dot{\hat{\mathbf{L}}}_N}\!\bigl( \omega, \hat{\mathbf{L}}_N \!\cdot\! \hat{\mathbf{S}}_1, \hat{\mathbf{L}}_N \!\cdot\! \hat{\mathbf{S}}_2, \hat{\mathbf{S}}_1, \hat{\mathbf{S}}_2 ; M, \eta, \chi_1, \chi_2 \bigr)
\times \hat{\mathbf{L}}_N \nonumber
\end{eqnarray}
(see Eqs.\ (2), (3), and (9) of Ref.~\cite{BuonannoChenVallisneri:2003b}).

The gravitational strain perturbation $h^{ij}$ is computed from the 
leading-order mass-quadrupole term specialized to circular orbits, 
following Finn and Chernoff~\cite{FinnChernoff:1993} 
(see also Sec.\  II C of Ref.~\cite{BuonannoChenVallisneri:2003b}). 
Since Finn and Chernoff use a fixed source coordinate system, the 
twice-differentiated mass-quadrupole tensor $Q^{ij}_c$ is a function of the 
orbital phase $\int \omega \, dt$ and of $\hat{\mathbf{L}}_N(t)$. The response 
of a ground-based interferometric detector is obtained by projecting $Q^{ij}_c$ 
onto a combination of unit vectors along the interferometer arms, which 
introduces a dependence on five angles that describe the relative direction 
($\Theta$ and $\varphi$, which subsumes the initial orbital phase of the binary) 
and orientation ($\theta$, $\phi$, and $\psi$) between the detector and the 
Finn--Chernoff source frame.

Equations (\ref{eq:targetomega}) and (\ref{eq:targetall}) are integrated 
numerically in the time domain until 
%$\omega$ reaches 
the minimum of the PN orbital energy 
$E_{3PN}(\omega,\hat{\mathbf{L}}_N,\hat{\mathbf{S}}_1, \hat{\mathbf{S}}_2,M,\eta,\chi_1,\chi_2)$ 
(see Eqs.\ (11) and (12) of Ref.~\cite{BuonannoChenVallisneri:2003b}) is reached or 
until $\dot{\omega}$ becomes negative. No attempt is made to describe the 
waveform beyond this stopping point, where it is assumed that the adiabatic 
approximation must break down. Altogether, the waveforms are functions of four 
mass and spin constants ($M$, $\eta$, $\chi_1$, and $\chi_2$), of six angles 
describing the orientations of $\hat{\mathbf{L}}_N$, $\hat{\mathbf{S}}_1$, and 
$\hat{\mathbf{S}}_2$ at a fiducial time and frequency, the five
direction and orientation angles and the distance of the detector from the source.
We note that the angles $\Theta$ and $\varphi$ are degenerate with the angles
given implicitly when we define $\hat{\mathbf{L}}_N$.
In this analysis we assume that the binary's orbits have become circularized (see brief
discussion in Sec.~\ref{evol}) and that the orbital eccentricity is zero.
Given this assumption we are able to describe the binary using 15 independent parameters.
 % Michele        
\rcsid$Id: detection_template_family.tex,v 1.28 2008/06/27 11:04:36 gareth Exp $
\section{Detection Template Family}
\label{sub:dtf}

As discussed in Sec.~\ref{evol},  when the binary components carry 
significant spins which are not aligned with the orbital angular momentum,  
spin-orbit and spin-spin couplings can induce a strong precession 
of the orbital plane, thus causing substantial modulation of the 
gravitational waves' amplitude and phase (see Fig.~\ref{waveforms}). 
Detection-efficient search templates must account for these 
effects of spin. A straightforward parametrization of search templates 
by the physical parameters that affect precession results in
very large template banks, which is computationally prohibitive. 
It is then necessary to reduce the number of waveform parameters
while still efficiently covering the parameter space of target waveforms. 

We shall denote by ``detection template family'' (DTF) a family of signals 
that captures the essential features of the true waveforms, but depend on a smaller 
number of parameters, either physical or phenomenological.  
At their best, DTFs can reduce computational 
requirements while achieving essentially the same detection performance 
as true templates. However, DTFs can include non-physical 
signal shapes that may increase the number of noise-induced triggers, 
affecting the upper-limit studies. Moreover, DTFs are also less 
adequate for parameter estimation, because the mapping between 
template and binary parameters is not one-to-one. 

In recent years several DTFs for precessing 
compact binaries have been proposed~\cite{Apostolatos:1994, Apostolatos:1995, Apostolatos:1996rg, GrandclementKalogeraVecchio:2003,
GrandclementKalogera:2003,grandclement:102002,BuonannoChenVallisneri:2003b, Buonanno:2005pt}. 
A DTF based on the so-called \emph{Apostolatos ansatz}~\cite{Apostolatos:1994, Apostolatos:1995} 
for the evolution of precession frequency was thoroughly investigated in 
Refs.~\cite{GrandclementKalogeraVecchio:2003,GrandclementKalogera:2003}. 
It was found that the computational requirements of the Apostolatos-type families 
are very high, and its signal-matching performances are not 
very satisfactory. An improved version using {\it spiky} templates 
was then proposed in Ref.~\cite{grandclement:102002}. 

After analyzing the physics of spinning-binary precession and waveform generation,   
the authors of Ref.~\cite{BuonannoChenVallisneri:2003b} showed that the 
modulational effects can be isolated in the evolution of the two gravitational wave 
polarizations (i.e., $h_+$ and $h_\times$), which combined with the detector's 
antenna patterns yield its response. 
As a result, the detector's response can be written as the product of a carrier
signal and a complex modulation factor, which can be handled using an extension 
of the Apostolatos ansatz. 
More explicitly, the modulated DTF in the frequency domain proposed 
in Ref.~\cite{BuonannoChenVallisneri:2003b} reads:
%\begin{eqnarray}
%&& h(\psi_{\mathrm{NM}},\mathcal{A}_{k},t_{0},\alpha _{k};f)=\left
%[\sum^{3}_{k=1}(\alpha_{k}+i\alpha _{k+3})\mathcal{A}_{k}(f)\right
%]\times \nonumber \\
%&& e^{2\pi ift_{0}}e^{i\psi_{\mathrm{NM}}(f)}\,\theta(f_{\mathrm{cut}} - f)  
%\quad (\mathrm{for}\;f>0)
%\label{DTF}
%\end{eqnarray}
\begin{eqnarray}
&& h(\psi_{\mathrm{NM}},t_{0},\alpha _{j};f)=\left
[\sum^{3}_{j=1}(\alpha_{j}+i\alpha _{j+3}) 
h_{j}(f)\right ]\times \nonumber \\
&& e^{2\pi ift_{0}} 
\theta(f_{\mathrm{cut}} - f)
\quad (\mathrm{for}\;f>0)
\label{DTF}
\end{eqnarray}
with $h(f)=h^{*}(-f)$ for $f<0$. 
The coefficients $\alpha_j$ 
in Eq.~(\ref{DTF}) are six real coefficients encoding the global 
phase, the strength of the amplitude modulation, its relative phase with respect to 
the leading order amplitude, and the internal (complex) phase of the modulations.
The coefficient $t_0$ is the time of arrival and $\theta(...)$ is the Heaviside step
function which is zero for all frequencies $f > f_{\rm cut}$. We use the parameter 
$f_{\rm cut}$ to terminate the template waveform once we believe it is no longer
an accurate representation of the true gravitational waveform (generally due 
to deviation away from the adiabatic approximation). 

In Eq.~(\ref{DTF})  
%the $\mathcal{A}_{j}(f)$ are the real amplitude functions:  
the functions $h_{j}(f) = \mathcal{A}_{j}(f) e^{i\psi_{\mathrm{NM}}(f)}$ 
are the {\it basis-templates} where $\mathcal{A}_{j}(f)$ are the real 
amplitude functions:  
\begin{eqnarray}
\label{newtonian} 
\mathcal{A}_{1}(f) &=& f^{-7/6}, \\
\label{sin}
\mathcal{A}_{2}(f) &=& f^{-7/6}\,\cos (\mathcal{B}), \\
\label{cos}
\mathcal{A}_{3}(f) &=& f^{-7/6}\,\sin (\mathcal{B}),
\end{eqnarray}
where $\mathcal{B} = \beta f^{-2/3}$ and $\beta$ 
is related to the frequency of precession \cite{Buonanno:2005pt} and
is used to capture the spin-induced modulation of the waveform. 
The function $\psi_\mathrm{NM}(f)$ represents the phase of the non-modulated 
carrier signal; it depends on the masses and spins of the binary's components 
and it can be computed in post-Newtonian (PN) theory. Here, as 
in Ref.~\cite{BuonannoChenVallisneri:2003b}, we express 
$\psi_\mathrm{NM}$ in terms of {\it only} two  
phenomenological parameters $\psi_0$ and $\psi_{3}$ 
\footnote{The symbol $\psi_{3}$ used here is equivalent to the $\psi_{3/2}$ used in the papers
by Buonanno, Chen and Vallisneri~\cite{BuonannoChenVallisneri:2003b}.}, i.e., 
\begin{equation}
\psi_{\mathrm{NM}}(f) = f^{-5/3} \,(\psi_{0} + \psi_{3} f )\,.
\end{equation}
In the case of single-spin binaries (i.e., only one of the bodies has spin), 
it is possible to (analytically) relate the three phenomenological 
parameters $\psi_{0}$, $\psi_{3}$ and $\beta$ 
with the four physical parameters $M$, 
$\eta$, $\kappa_1$ and $\chi_1$~\cite{Buonanno:2005pt}. 
The physical parameter $\kappa$ is the cosine of the angle between the direction
of the (total) spin and the orbital angular momentum and in this case would
be $\kappa_{1} \equiv \hat{\mathbf{L}}_N \cdotp \hat{\mathbf{S}}_1$.
However, 
for double-spin binaries --- which is the case investigated 
in this paper --- the mapping is not analytical and the number 
of physical parameters is greater than four, resulting in an intractably 
large template bank. Within the spirit of DTF and, as a first 
step in implementing search templates for spinning, precessing binaries, 
we proceed here with the three phenomenological parameters $\psi_{0}$, $\psi_{3}$ 
and $\beta$. 

The DTF described by Eq.~(\ref{DTF}) generalizes the Apostolatos ansatz in two ways: 
it allows a {\it complex} phase offset between i) the leading order 
%Newtonian 
$f^{-7/6}$ 
amplitude term (Eq.~\ref{newtonian}) and the sinusoidal amplitude terms 
(Eqs.~\ref{sin} and \ref{cos}) and 
ii) between the cosine and sine modulation terms. 
Quite interestingly, as shown in Ref.~\cite{Buonanno:2005pt}, 
by an appropriate choice of the phenomenological coefficients $\alpha_{1 \cdots 6}$,  
the DTF also has the ability to generate higher harmonics 
which arise in the target signal discussed in Sec.~\ref{sub:target}. Those higher 
harmonics are caused by oscillations in the components of the gravitational wave 
polarization tensor 
and not directly by the precession of the orbital angular momentum and spins, 
and should be reproduced by the search templates in order not to lose efficiency. 

Henceforth, we will treat $\psi_{0}$, $\psi_{3}$ and $\beta$ as {\it intrinsic} 
parameters and the $\alpha_{1 \cdots 6}$ and $t_0$ as {\it extrinsic} parameters.
%Intrinsic parameters describe the source itself (e.g., masses, spins)
%and require the use of templates in order to find their value.
%On the other hand, extrinsic parameters relate to the observer's relation to 
%the source (e.g., distance of the source from the observer, the amplitude and time of arrival 
%of the gravitational wave at the observer) and can be found simply through maximization 
%of the SNR with respect to the extrinsic parameter 
%(e.g., measurement of a signal's time of arrival using an FFT).
Intrinsic parameters describe the source itself (e.g., masses, spins).
To maximize the SNR with respect to the intrinsic parameters we must construct
templates corresponding to different values of the intrinsic parameters
and measure the SNR obtained by each of these templates with our detector data.
On the other hand, extrinsic parameters describe the observer's relation to 
the source (e.g., distance of the source from the observer, 
the amplitude and time of arrival of the gravitational wave at the observer).
Maximization of the SNR with respect to extrinsic parameters can be performed
automatically (e.g., measurement of a signal's time of arrival using an FFT) and
is computationally cheaper than maximization of the SNR with respect to the intrinsic
parameters.
 
In practice we set $f_{\rm cut}$ to the frequency of the gravitational wave emission at
the last stable orbit (LSO) which we estimate using
\begin{equation}
%f_{\rm cut} = (-16 \pi \psi_0)/(\psi_3 r_{\rm LSO}^{3/2})
f_{\rm cut} \approx f_{\rm LSO} = \frac{M^{1/2}} {\pi r_{\rm LSO}^{3/2}} 
\label{fcut}
\end{equation}
where $r_{\rm LSO} = 6M$ is the separation of the binary's components and
the total mass $M$ is estimated from $\psi_{0}$ and $\psi_{3}$ using approximate
relationships between phenomenological and physical parameters we introduce in the
next section, see Eqs.~(\ref{psi0m},\ref{psi3m}).

To assess whether a stretch of detector data contains a gravitational wave signal we calculate the 
signal-to-noise ratio (SNR) which is the cross-correlation of our templates with the data. 
The full process of deciding whether a detection has been made is described in Sec.~\ref{sub:pipeline} 
of this paper and more fully in the companion papers~\cite{findchirppaper,LIGOS3S4all}.
We can simplify the calculation of SNR by orthonormalization
of the amplitude functions $\mathcal{A}_{k}$. 
We obtain the orthonormalized amplitude functions, denoted $\mathcal{\widehat{A}}_{k}$, using the 
Gram-Schmidt procedure which leads to the transformations:
\begin{eqnarray}
\mathcal{A}_1 & \to & \mathcal{\widehat{A}}_1 = \frac {\mathcal{A}_1}{||\mathcal{A}_{1}||^{1/2}} \nonumber \\
\mathcal{A}_2 & \to & \mathcal{\widehat{A}}_2
 =  \frac {\mathcal{A}_2 - \left < \mathcal{A}_2 , \mathcal{\widehat{A}}_1 \right > \mathcal{\widehat{A}}_1 }
 {|| \mathcal{A}_2 - \left < \mathcal{A}_2 , \mathcal{\widehat{A}}_1 \right > \mathcal{\widehat{A}}_1 || ^{1/2} } \\
\mathcal{A}_3 & \to & \mathcal{\widehat{A}}_3 = \frac {\mathcal{A}_3 - \left < \mathcal{A}_3 , \mathcal{\widehat{A}}_1 \right > \mathcal{\widehat{A}}_1 - \left < \mathcal{A}_3 , \mathcal{\widehat{A}}_2 \right > \mathcal{\widehat{A}}_2 }
%{\mathcal{N}_3}
{|| \mathcal{A}_3 - \left < \mathcal{A}_3 , \mathcal{\widehat{A}}_1 \right > \mathcal{\widehat{A}}_1 - \left < \mathcal{A}_3 , \mathcal{\widehat{A}}_2 \right > \mathcal{\widehat{A}}_2 || } \nonumber
\end{eqnarray}
where we use $||a||$ to represent the inner product of a function with itself: $||a|| = \left < a, a \right >$.
Throughout we will use the real-valued inner product:
\begin{equation}
\label{innerproduct}
\left <a,b \right > = 4 \Re \int_0^\infty df {\tilde{a}^*(f) \tilde{b}(f) \over
S_h(f)}
\end{equation}
where $S_h (f)$ is an estimate of the noise power spectral density of the data.
%where $\left <a,a \right > = 1$ and $\left <a,b \right > = 0$ when $a$ and $b$ are orthonormal.
The final form of the orthonormalized amplitude functions are very long and for that reason not reproduced here.
The DTF in terms of the orthonormalized amplitude functions has the exact same form as that shown in 
Eq.~(\ref{DTF}) with $h$, $h_{j}$ and $\alpha_{j}$ 
replaced by $\hat{h}$, $\hat{h}_{j}$ and $\hat{\alpha}_{j}$ respectively.
Demanding templates normalized so that $\left< h,h \right> = \left< \hat{h}, \hat{h} \right> = 1$ leads to the
constraint $\sum_{j=1}^{6} \hat{\alpha}_j^2 = 1$.
Having defined the orthonormalized amplitude functions $\mathcal{\widehat{A}}_{k}$ we can calculate the SNR, $\rho$:
\begin{equation}
\label{defrho}
\rho  =  \max_{t_0, \alpha_j} \left < x,h(t_0 , \alpha_j ) \right > =
\max_{t_0}
 \sqrt{ \sum_{j=1}^{6}  \left < x,\hat{h}_j(t_0) \right > ^2 },
\end{equation}
where $x$ is the detector data and the orthonormalized basis-templates are given by
\begin{eqnarray} 
\hat{h}_{j} &=&    \mathcal{\widehat{A}}_j(f) e^{i \psi_{\mathrm{NM}} (f)} \quad \mathrm{for} \; j=1,2,3 \; \mathrm{and} \nonumber \\
\hat{h}_{j} &=&  i \mathcal{\widehat{A}}_{j-3}(f) e^{i \psi_{\mathrm{NM}} (f)} \quad \mathrm{for} \; j=4,5,6.
\end{eqnarray}
%In Eq.~(\ref{defrho}), $x$ represents the detector data and in Eq.~(\ref{innerproduct}) $S_{h}(f)$ represents 
%the noise power spectral density of this detector data.
Note that we do not explicitly need to calculate $\alpha_{1 \cdots 6}$ in order to calculate the SNR but that 
they can be found simply if required:
$\hat{\alpha}_{j} = \left < x,\hat{h}_j(t_0) \right > / \rho$.

For Gaussian white noise, $\rho^2$ will, in general, have a $\chi^2$ distribution with 6 degrees of
freedom. In the case where the spin parameter $\beta = 0$ we find that $\mathcal{\widehat{A}}_2$ and $\mathcal{\widehat{A}}_3$ 
both vanish and that $\rho^2$ is described by a $\chi^2$ distribution with 2 degrees of freedom.
To reflect the increased freedom we choose a higher SNR threshold, $\rho_{*} = 12$ when $\beta \neq 0$ and a lower value of
%$\rho_{*} = \sqrt{125}$
$\rho_{*} \approx 11.2$ when $\beta = 0$. 
These values were chosen to give approximately the same number of triggers when analyzing
Gaussian white noise and to ensure that the number of triggers produced during the real search was manageable.
 % Alessandra
\rcsid$Id: template_bank.tex,v 1.33 2008/01/16 11:09:16 gareth Exp $
\section{Template Bank}
\label{sub:templatebank}

%We now consider the problem of placing templates in the 3-dimensional
%parameter space defined by our intrinsic parameters $\psi_0$, $\psi_3$
%and $\beta$. We calculate the metric of the intrinsic parameter space
%which enables us to calculate the mismatch or loss of SNR between
%points in this space.
%In order to detect the presence of gravitational wave signals we will
%filter our detector data against a bank of templates and record
%the SNRs obtained. 
Since we will not know the parameters describing an incident gravitational waveform
a priori, we must filter our detector data with a set of templates known
as a {\it template bank}.
Neglecting the effects of noise, we would expect that the template yielding the largest
SNR to be the best representation of an incoming signal.
Due to the discrete nature of the template bank (it must be discrete since it can only
contain a finite number of templates) we will lose SNR due to mismatches between the
intrinsic parameters of any gravitational wave signal and the best template.
By placing templates with an appropriate density, we can limit the maximum mismatch
between signal and template intrinsic parameters and hence limit the loss of SNR
caused by the discreteness of the bank.
The spacing of templates in the intrinsic parameter space required to limit this 
mismatch can be found using the metric on the signal manifold \cite{Owen:1995tm,Owen:1998dk}.
%In this section we describe the design and testing of the template bank.
%The largest challenge when designing this bank was to find a suitable 
%approximation for the parameter space metric, which in general can be quite complicated.
In this section we describe the calculation of the metric, the template placement algorithm 
and comparisons with other banks before discussing the testing of the bank 
using software-injected simulated signals.

\subsection{Metric calculation}
%In order to choose the sparsest, and therefore most efficient, spacing
%of templates that will still effectively capture a variety of target waveforms
%we make use the metric of our intrinsic parameter space.
%This parameter space metric describes the matched-filter overlap between
%two waveforms, each generated with a given set of intrinsic parameters 
%(i.e., $\psi_0$, $\psi_3$ and $\beta$).

In this search we use a simple metric based on the strong modulation
approximation described below.
The rationale is that systems with waveforms only weakly modulated by
spin-induced precession should be detectable with high efficiency by a 
non-spinning binary search, e.g.,~\cite{LIGOS3S4all}.
Thus we concentrate on designing a bank that will capture systems whose
waveforms will be strongly modulated.
The metric calculation and template placement (or tiling) algorithms
become much simpler in the strong modulation limit.
More recently, more precise treatments of the full metric on the DTF
parameter space have become available~\cite{Pan:2003qt, Buonanno:2005pt}
and work is in progress to incorporate them into future searches.

In the strong modulation approximation, the orbital plane is assumed to precess 
many times as the gravitational wave sweeps through the LIGO band of good sensitivity.
Also the opening angle between the orbital and spin angular momentum is 
assumed to be large, corresponding to large amplitude modulations of
the signal.
Mathematically this corresponds to the statement that the precession phase
$\mathcal{B}$ sweeps through many times $2\pi$ and thus that the
basis-templates $h_j$ are nearly
orthonormal (without need for the Gram-Schmidt procedure).
Below we shall see that this assumption places a condition on the precession parameter
$\beta$, which for the initial LIGO design noise power spectral 
density~\cite{Abramovici:1992ah}
%\cite{Damour:2000zb} 
corresponds to $\beta \gtrsim 200$~Hz$^{2/3}$.

We can relate this condition for validity of the strong modulation 
approximation to the astrophysical parameters of system. 
Na\"{i}vely we can put the phenomenological parameters in terms of astrophysical
parameters using:
\begin{eqnarray}
\label{psi0m}
\psi_{0} &=& \frac{3}{128} [\pi (m_1+m_2)]^{-5/3} \frac{(m_1+m_2)^2} {m_1m_2},
\\
\label{psi3m}
\psi_{3} &=& -\frac{3\pi}{8} [\pi (m_1+m_2)]^{-2/3} \frac{(m_1+m_2)^2}
{m_1m_2},
\\
\label{betam}
\beta &=& 258\text{ Hz}^{2/3} \left(1 + \frac{3m_2}{4m_1} \right)
\frac{m_1}{m_2} \chi \left( \frac{M_\odot} {m_1+m_2} \right)^{2/3}
\end{eqnarray}
acknowledging that, in reality, the true signal manifold and phenomenological
template manifold do not map this simply.
The equations for $\psi_{0}$ and $\psi_{3}$ can be found by considering the
expansion for the gravitational wave phase $\psi(f)$ 
given in terms of masses (e.g., Eqs.~(3.3) and (3.4) of \cite{arun:084008}) 
and equating the dominant terms of this expansion to those with the same 
frequency exponent in the expansion for gravitational wave phase given in terms
of $\psi_{0,1,...}$ in \cite{BuonannoChenVallisneri:2003a, BuonannoChenVallisneri:2003b}.
The effects of spin are neglected in these approximations of $\psi_{0}$ and $\psi_{3}$.
The equation for $\beta$ arises by recognizing that $\beta$ is related to the evolution
of the rate of precession, see Eq.~(45) of \cite{Apostolatos:1994} and \cite{Buonanno:2005pt}
for further discussion \footnote{Please note that a small error appears in Eq.~(45) of \cite{Apostolatos:1994}.
The terms $1 + 3 m_{1}/ 4 m_{2}$ should in fact read $1 + 3 m_{2}/ 4 m_{1}$. 
Corrected forms of these expressions occur in Eq.~(29) of \cite{Apostolatos:1995}}.

The constraint for validity of the strong modulation approximation is that the mass ratio 
%must exceed approximately 2:1.
must satisfy $m_{2} / m_{1} \gtrsim 2$.
Also, we specify that the total mass be less than some value 
%(here about 15~$M_\odot$) 
(here  $\sim 15~M_\odot$) 
so that the waveforms do not begin far enough into the
non-linear region to require extra phenomenological parameters.
Thus the parameter space region of such a search may be expressed solely in
terms of the range of masses for the lower-mass body.
%In this search the range used was 
%$1.0~M_\odot < m_{1} < 3.0~M_\odot$, 
%a likely range of masses for neutron stars, corresponding formally to a 
%$6.0~M_\odot < m_{2} < 12.0~M_\odot$ 
%range for the more massive body.
In this search the range used for $m_{1}$ was
$1.0~M_{\odot} < m_{1} < 3.0~M_{\odot}$, 
a likely range of masses for neutron stars, and 
$6.0~M_{\odot} < m_{2} < 12.0~M_{\odot}$ was used as the
range for the more massive body.
Thus, astrophysically this search is directed at NS-BH
systems or BH-BH systems with unequal masses.
These mass ranges are converted into ranges of $\psi_{0}$
and $\psi_{3}$ using Eqs.~(\ref{psi0m}) and (\ref{psi3m})
which define the region of parameter space we populate with
templates.
Due to the inexact nature of these equations
we know that the range of masses for which the template bank
obtains its highest matches may differ from the range of masses
we use to specify the region of $(\psi_{0}, \psi_{3})$.
We will show in Sec.~\ref{sub:testingbank} that a template bank 
generated using the range of masses just specified
%$1.0~M_{\odot} < m_{1} < 3.0~M_{\odot}$ and 
%$6.0~M_{\odot} < m_{2} < 12.0~M_{\odot}$
yields high matches (greater than $0.9$) for binaries with physical
masses in the asymmetric range $1.0~M_{\odot} < m_{1} < 3.0~M_{\odot}$ and
$12.0~M_{\odot} < m_{2} < 20.0~M_{\odot}$.
We shall also show that this search is efficient for non-spinning 
systems as well as for spinning ones.

We derive the metric components in the manner of Ref.~\cite{Owen:1995tm}.
Starting from the detection statistic $\rho^2$ (the square of
Eq.~(\ref{defrho})), let us take our data $x$ to have the form of a template 
with parameters slightly perturbed from those of the template $h$ we filter it with:
\begin{eqnarray}
\tilde{x}(f) &=& (\alpha_1 + i\alpha_2) e^{i(\psi_{\mathrm{NM}} + d\psi_{\mathrm{NM}})} \nonumber
\\ 
&&  + (\alpha_3 + i\alpha_4) \cos(\mathcal{B} + d\mathcal{B}) 
e^{i(\psi_{\mathrm{NM}} + d\psi_{\mathrm{NM}})} \nonumber
\\ 
&& + (\alpha_5 + i\alpha_6) \sin(\mathcal{B} +
d\mathcal{B}) e^{i(\psi_{\mathrm{NM}} + d\psi_{\mathrm{NM}})} 
\end{eqnarray}
Note that only the intrinsic parameters are perturbed, as the maximization
takes care of the extrinsic parameters.
Expanding to second order in the perturbation, we have
\begin{eqnarray}
\tilde{x}(f) &\approx& \left( 1 + i\,d\psi_{\mathrm{NM}} - {1\over2} d\psi_{\mathrm{NM}}^2 \right)
\bigg\{ (\alpha_1 + i\alpha_2) {h}_1
\nonumber
\\
&& + (\alpha_3 + i\alpha_4) \left[ \left( 1 - {1\over2} d\mathcal{B}^2
\right) {h}_2 - d\mathcal{B} {h}_3 \right] \nonumber
\\
&&  + (\alpha_5 + i\alpha_6) \left[ \left( 1 - {1\over2}
d\mathcal{B}^2 \right) {h}_3 + d\mathcal{B} {h}_2 \right] \bigg\}.
\end{eqnarray}
Under the approximation that the $h_j$ are orthonormal, we get
\begin{eqnarray}
\left < x,h_1 \right > &=& \alpha_1 \left[ 1 - {1\over2} F\left(
d\psi_{\mathrm{NM}}^2 \right) \right] - \alpha_2 F(d\psi_{\mathrm{NM}}), \nonumber
\\
\left < x,h_4 \right > &=& \alpha_2 \left[ 1 - {1\over2} F\left(
d\psi_{\mathrm{NM}}^2 \right) \right] + \alpha_1 F(d\psi_{\mathrm{NM}}), \nonumber
\\
\left <x,h_2 \right > &=& \alpha_3 \left[ 1 - {1\over2} F\left(
d\psi_{\mathrm{NM}}^2 \right) - {1\over2} F\left( d\mathcal{B}^2 \right) \right]
\nonumber
\\
&& - \alpha_4 F(d\psi_{\mathrm{NM}})
+ \alpha_5 F(d\mathcal{B}) - \alpha_6 F(d\psi_{\mathrm{NM}}\, d\mathcal{B}), \nonumber
\\
\left < x,h_5 \right > &=& \alpha_4 \left[ 1 - {1\over2} F\left( d\psi_{\mathrm{NM}}^2 \right) -
{1\over2} F\left( d\mathcal{B}^2 \right) \right] + \alpha_3 F(d\psi_{\mathrm{NM}})
\nonumber
\\
&& + \alpha_6 F(d\mathcal{B}) + \alpha_5 F(d\psi_{\mathrm{NM}}\, d\mathcal{B}), \nonumber
\\
\left < x,h_3 \right > &=& \alpha_5 \left[ 1 - {1\over2} F\left( d\psi_{\mathrm{NM}}^2 \right) -
{1\over2} F\left( d\mathcal{B}^2 \right) \right] - \alpha_6 F(d\psi_{\mathrm{NM}})
\nonumber
\\
&& - \alpha_3 F(d\mathcal{B}) + \alpha_4 F(d\psi_{\mathrm{NM}}\, d\mathcal{B}), \nonumber
\\
\left < x,h_6 \right > &=& \alpha_6 \left[ 1 - {1\over2} F\left( d\psi_{\mathrm{NM}}^2 \right) -
{1\over2} F\left( d\mathcal{B}^2 \right) \right] + \alpha_5 F(d\psi_{\mathrm{NM}})
\nonumber
\\
&& - \alpha_4 F(d\mathcal{B}) - \alpha_3 F(d\psi_{\mathrm{NM}}\, d\mathcal{B}),
\label{xhrels}
\end{eqnarray}
%where $F$ indicates the functional referred to as $\mathcal{J}$
%in Ref.~\cite{Owen:1995tm}.
where $F$ is a functional (originally defined in Ref.~\cite{Owen:1995tm} as
$\mathcal{J}$) given by
\begin{equation}
F(a) = \frac{1}{I_{7}} \int_{f_{\rm min}/f_{0}}^{f_{\rm max}/f_{0}} dx 
\frac
{x^{-7/3}}
{S_{h}(x f_{0})} a(x)
\end{equation}
and the noise moment $I$ is itself defined as
\begin{equation}
I_{q} \equiv \int_{f_{\rm min}/f_{0}}^{f_{\rm max}/f_{0}} dx 
\frac
{x^{-q/3}}
{S_{h}(x f_{0})}
\label{funI}
\end{equation}
where $f_{\rm min}$ and $f_{\rm max}$ define the range of frequencies we integrate over.
In S3 we used a lower cutoff frequency of $70$~Hz, chosen to exclude lower
frequencies for which the detector's power spectral density was significantly
non-stationary, and an upper frequency
corresponding to the Nyquist frequency, in this case $1024$~Hz.
Inserting the relations from Eq.~(\ref{xhrels}) into Eq.~(\ref{defrho}) and keeping up to second
order perturbations, we obtain
\begin{eqnarray}
\label{thingy}
\sum_{j=1}^6 \left< x, h_j \right>^2 &=& \sum_{j=1}^6 \alpha_j^2 \left[ 1
- F(d\psi_{\mathrm{NM}}^2) + F(d\psi_{\mathrm{NM}})^2 \right] \nonumber
\\
&-& \sum_{j=3}^6 \alpha_j^2 \left[ F(d\mathcal{B}^2) - F(d\mathcal{B})^2
\right] \nonumber 
\\
&-& \bigg[ 2\left( \alpha_3 \alpha_6 \right. - \left. \alpha_4 \alpha_5 \right) 
\nonumber \\
&\times& \left[ F(d\psi_{\mathrm{NM}}\, d\mathcal{B}) -
F(d\psi_{\mathrm{NM}}) F(d\mathcal{B}) \right] \bigg].
\end{eqnarray}

To finish computing the perturbed $\rho^2$ we must maximize Eq.~(\ref{thingy})
over the coalescence time and $\alpha_j$ (subject to the constraint
$\sum_{j=1}^6 \alpha_j^2 = 1$ since we are dealing with normalized
waveforms).
Maximization over $\alpha_j$ is performed straightforwardly using Lagrange multipliers.
We find $\alpha_1 = \alpha_2 = 0$, $\alpha_3 = -\alpha_6$, and $\alpha_4 =
\alpha_5$, which leads to
\begin{eqnarray}
\max_{\alpha_j} && \left< x,h_j \right>^2 = 1 - F(d\psi_{\mathrm{NM}}^2) +
F(d\psi_{\mathrm{NM}})^2 - F(d\mathcal{B}^2) \nonumber
\\
&& + F(d\mathcal{B})^2 + F(d\psi_{\mathrm{NM}}\, d\mathcal{B}) - F(d\psi_{\mathrm{NM}})
F(d\mathcal{B}).
\end{eqnarray}
We incorporate the time-dependence of $\rho^{2}$ into the template's phasing 
and expand the phase functions in terms of the phenomenological parameters
and coalescence time $t_{c}$
\begin{eqnarray}
d\psi_{\mathrm{NM}} &=& d\psi_{0}\, f^{-5/3} + d\psi_{3}\, f^{-2/3} + 
2 \pi f dt_{c},
\\
d\mathcal{B} &=& d\beta\, f^{-2/3}.
\end{eqnarray}
Using the definition of the metric~\cite{Owen:1995tm} to write
\begin{equation}
\rho^2 = 1 - 2g_{ab} d\lambda^a d\lambda^b,
\end{equation}
we obtain the metric components
\begin{eqnarray}
\label{gnoproj}
2 g_{t_c t_c} &=& 4 \pi^{2} \left( J_1 - J_4^2 \right), \nonumber
\\
2 g_{t_c \psi_0} &=& 2 \pi \left( J_9 - J_4J_{12} \right), \nonumber
\\
2 g_{t_c \psi_3} &=& 2 \pi \left( J_6 - J_4J_9 \right), \nonumber
\\
2 g_{t_c \beta} &=& (- \pi /2) \left( J_6 - J_4J_9 \right), \nonumber
\\
2 g_{\psi_{0} \psi_{0}} &=& J_{17} - J_{12}^2, \nonumber
\\
2 g_{\psi_{0} \psi_{3}} &=& J_{14} - J_9J_{12}, \nonumber
\\
2 g_{\psi_{0} \beta} &=& (-1/2) \left( J_{14} - J_9J_{12} \right), \nonumber
\\
2 g_{\psi_{3} \psi_{3}} &=& J_{11} - J_9^2, \nonumber
\\
2 g_{\psi_{3} \beta} &=& (-1/2) \left( J_{11} - J_9^2 \right), \nonumber
\\
2 g_{\beta \beta} &=& J_{11} - J_9^2
\end{eqnarray}
%
%\begin{eqnarray}
%\label{gnoproj}
%2\gamma_{t_c t_c} &=& 4 \pi^{2} \left( J_1 - J_4^2 \right), \nonumber
%\\
%2\gamma_{t_c \psi_0} &=& 2 \pi \left( J_9 - J_4J_{12} \right), \nonumber
%\\
%2\gamma_{t_c \psi_3} &=& 2 \pi \left( J_6 - J_4J_9 \right), \nonumber
%\\
%2\gamma_{t_c \beta} &=& (-1/2) \left( J_6 - J_4J_9 \right), \nonumber
%\\
%2\gamma_{\psi_{0} \psi_{0}} &=& J_{17} - J_{12}^2, \nonumber
%\\
%2\gamma_{\psi_{0} \psi_{3}} &=& J_{14} - J_9J_{12}, \nonumber
%\\
%2\gamma_{\psi_{0} \beta} &=& (-1/2) \left( J_{14} - J_9J_{12} \right), \nonumber
%\\
%2\gamma_{\psi_{3} \psi_{3}} &=& J_{11} - J_9^2, \nonumber
%\\
%2\gamma_{\psi_{3} \beta} &=& (-1/2) \left( J_{11} - J_9^2 \right), \nonumber
%\\
%2\gamma_{\beta \beta} &=& J_{11} - J_9^2
%\end{eqnarray}
before projecting out the coalescence time $t_c$.
Here we have used $J_{q}$ to represent the normalized noise moments given
by~\cite{Poisson:1995ef}
\begin{equation}
J_q \equiv I_q/I_7
\end{equation}
where the noise moment $I$ was defined in Eq.~(\ref{funI}).
%where the $I_{p}$ represent the non-normalized moments
%\begin{equation}
%I_p = \int_0^\infty df \left[ f^{p/3} S_h(f) \right]^{-1}.
%\end{equation}
These moments give us a way of checking when the strong modulation
approximation is valid.

\begin{figure}[hbtp]
\begin{center}
\includegraphics[width=0.5\textwidth]{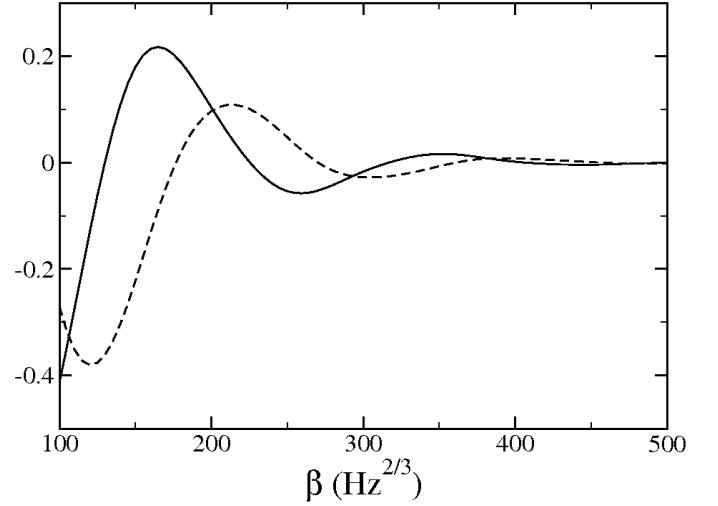}
\end{center}
\caption{Moment functions $C_7(\beta)$ (solid line) and $S_7(\beta)$
(dashed line) for the initial LIGO design noise power spectral density.
For values of $\beta \gtrsim 200$~Hz$^{2/3}$ we see that these moments become
small and can be neglected - this is what we call the {\it strong modulation approximation}.
%Some of the fluctuations in these moments occurring for $\beta>100$~Hz$^{2/3}$ are due to
%the probabilistic numerical estimates used.
}
\label{momentsplot}
\end{figure}

If we had not made the strong modulation approximation, we would also need the
functions
\begin{eqnarray}
C_p(\beta) &=& \int_0^\infty df\, \left[ f^{p/3} S_h(f) \right]^{-1} \cos
\mathcal{B}(f) /I_7,
\\
S_p(\beta) &=& \int_0^\infty df\, \left[ f^{p/3} S_h(f) \right]^{-1} \sin
\mathcal{B}(f) /I_7,
\end{eqnarray}
which we call the cosine and sine moment functions.
The inner products of the basis templates $h_j$ with each other (prior to the Gram-Schmidt
procedure) are proportional to these moment functions, and thus the strong
modulation approximation corresponds to assuming that $C_7$ and $S_7$ are small
compared to unity.
For the initial LIGO design noise power spectral density curve \cite{LIGO-E950018-02-E} 
the moment functions are plotted
in Fig.~\ref{momentsplot}.
We see that the strong modulation approximation should hold (to about the
10\% level) for $\beta \gtrsim 200$~Hz$^{2/3}$. 
See also Fig.~15 of Ref.~\cite{Buonanno:2005pt}, discussed more below, which
shows approximately the same behavior.

After projecting the coalescence time out of Eq.~(\ref{gnoproj}) and
dropping $\psi\beta$ cross terms (which simplifies the template placement and changes the
volume per template by less than 3\%), we obtain
\begin{eqnarray}
\label{metric}
2g_{\psi_{0} \psi_{0}} &=& J_{17} - J_{12}^2 - \left( J_9 - J_4J_{12} \right)^2 /
\left( J_1 - J_4^2 \right), \nonumber
\\
2g_{\psi_{0} \psi_{3}} &=& J_{14} - J_9J_{12} - (J_6 - J_4J_9) (J_9 -
J_4J_{12}) \nonumber
\\
&& / \left( J_1 - J_4^2 \right), \nonumber
\\
2g_{\psi_{0} \beta} &=& 0, \nonumber
\\
2g_{\psi_{3} \psi_{3}} &=& J_{11} - J_9^2 - (J_6 - J_4J_9)^2 / \left( J_1 - J_4^2
\right), \nonumber
\\
2g_{\psi_{3} \beta} &=& 0, \nonumber
\\
2g_{\beta \beta} &=& J_{11} - J_9^2 - (J_6 - J_4J_9)^2 / 4\left( J_1 - J_4^2
\right).
\end{eqnarray}

\subsection{Template placement algorithm}

We set the density of our template bank in terms of the {\it minimal match}
($MM$), defined to be the lowest match that can be obtained between a signal
and the nearest template~\cite{Owen:1995tm}.
A template bank designed to have minimal match $MM=0.95$ would therefore
suffer no more than a $1-MM = 5\%$ loss in SNR due to mismatch between
the parameters of a signal and the best possible template in the bank 
(assuming that the signal and templates are from the same family).

The metric components shown in Eq.~(\ref{metric}) are constant in the
strong modulation approximation, which enables us to use a simple template
placement algorithm.
We use a body-centred cubic (BCC) lattice which is the most efficient
template placement in three dimensions.
We first diagonalize the metric, which leaves the $\beta$ parameter
unchanged but gives us new ``horizontal'' parameters $\psi_{0}'$ and
$\psi_{3}'$.
Starting on the plane $\beta=0$, we draw a box in the primed coordinates
which encloses the part of that plane to be searched.
Beginning at one corner of this box, we step in the primed ``horizontal''
coordinates by amounts $(4/3)\sqrt{2(1-MM)/E}$, where $E$ is the
corresponding eigenvalue of the metric, i.e., 
$g_{\psi_{0}' \psi_{0}'}$ or 
$g_{\psi_{3}' \psi_{3}'}$.
At each point we transform to the mass parameters using Eqs.~(\ref{psi0m}) and 
(\ref{psi3m}) and check if we are in the targeted region of physical mass
space.
If the point is within that region, we add a template to the list.
Once a plane of constant $\beta$ is filled, we move ``up'' a distance 
in $\beta$ equal to $(2/3)\sqrt{2(1-MM)/g_{\beta\beta}}$, and 
lay a ``horizontal'' grid which is staggered half a cell (in both primed
directions) from the previous one.
Thus a BCC lattice is formed.

Such a simple template placement algorithm is susceptible to the ``ragged edges'' problem.
That is, there will be some areas near the edge of the targeted region of
parameter space that will match the nearest template at a level less than
$MM$.
The problem appears in other template placement algorithms such as those of
Refs.~\cite{Owen:1998dk, BBCCS:2006}, and sometimes is addressed in a
complicated way.
Our solution is simple and practical. 
In stepping around the $(\psi_0', \psi_3')$ plane, we check to see if we
have crossed the edge of the targeted region.
If we find ourselves at a point outside of the targeted region, we check to see 
whether the point halfway between the current position and the previously laid 
template is itself within the targeted region. If so, we add a template there.
%If we find a template outside of the targeted region, we check to see 
%whether a point halfway towards the edge is itself within
%the targeted region. If so, we add a template there.
Although the edges of the targeted region are curved, the radius of 
curvature is many template spacings meaning that we can treat the 
edges as fairly straight. 
This simple method solves the
ragged edges problem while resulting in a small number of additional
templates.

As mentioned earlier, we choose $f_{\rm cut}$, the frequency at which we end 
our template, to be the frequency of gravitational wave emission at the
last stable orbit. However, we compute metric components by effectively taking 
$f_{\rm cut}$ to infinity, which gains us simplicity at the cost of a small over-coverage.

We can compare the simplified template bank used here to those proposed in the
literature, particularly in Refs.~\cite{Pan:2003qt, Buonanno:2005pt}.
Although neither of those articles actually constructs a template bank or gives
explicit metric components, we can find a point of comparison.
Figure~15 of Ref.~\cite{Buonanno:2005pt} plots the coordinate volume per
template as a function of $\beta$, assuming a simple cubic lattice with
$MM=0.97$ and an analytical approximation to the initial LIGO noise curve.
In the high-$\beta$ (strong modulation) limit, their volume tends to
$\sim 5\times10^6$~Hz$^3$.
For the same $MM$, lattice, and noise curve, our volume per template is
$\sim 6.4\times10^6$~Hz$^3$.
Thus, our grid is slightly sparser than that of Ref.~\cite{Buonanno:2005pt}.
Most of the difference is because they define their final metric (on the space of
intrinsic parameters only) in terms of a ``minimax'' overlap, which is more
restrictive than the metric described here.
The issue is that the spacing on the intrinsic parameter space in general
depends on the extrinsic parameters, and there are multiple ways to remove this
dependence.
The minimax criterion of Ref.~\cite{Buonanno:2005pt} assumes the worst case (in
terms of extrinsic parameters) or tightest spacing for each point in parameter
space, and thus is tighter (lower template volume) than it needs to be.
Spin-induced precession of the orbital plane will cause sidebands either side
of the carrier frequency.
The metric we describe is constructed implicitly assuming that there is always non-zero
power at the carrier frequency and both precession sidebands, which eliminates
a set of measure zero of worst-case points in the extrinsic parameter space.
The template bank tests described below verify that the loss of efficiency due
to neglecting the worst-case extrinsic parameters is no more than a few
percent.

\begin{figure}
\begin{center}
\includegraphics[width=0.5\textwidth]{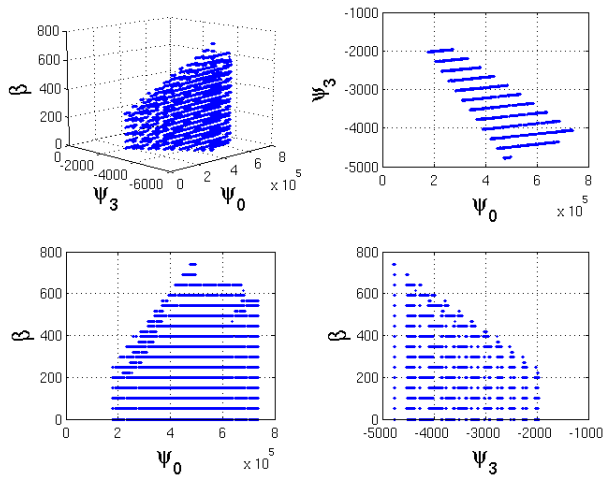}
\end{center}
\caption{A template bank generated with minimal match $= 0.95$ 
%using the initial LIGO design noise power spectral density. 
using 2048 seconds of H1 data taken during S3.
The crosses show the positions of individual templates in the 
$(\psi_0,\psi_3,\beta)$ parameter space. For each template a value 
for the cutoff frequency $f_{\rm cut}$ is estimated using Eq.~\ref{fcut}.
This bank requires a 3-dimensional template placement scheme in order
to place templates in the $(\psi_0,\psi_3,\beta)$ parameter space. Previous
searches for non-spinning systems have used 2-dimensional placement schemes.}
\label{templatebank}
\end{figure}

For the real S3 noise spectra which were used to construct the template banks
in this search, template numbers were typically $2-6 \times 10^3$ 
in H1 and L1 when prescribing a minimal match of $0.95$. The number of templates 
was larger in H2 compared to the other detectors and also increased with time
to $\sim 1.6 \times 10^4$ towards the end of S3 due to a flattening of the noise power
spectrum in H2.
Although a minimal match of $0.95$ was prescribed the effective minimal match
of the template banks generated was reduced to $\sim 0.93$ due to a small
calculation error.
%Figure \ref{templatebank} shows a template bank generated using 
%the initial LIGO design noise power spectral density with a minimal match $=0.95$.
Figure \ref{templatebank} shows a template bank generated using 
2048 seconds of H1 data and with a prescribed minimal match of $0.95$.

 % Ben
\rcsid$Id: testing_bank.tex,v 1.10 2008/06/27 11:04:37 gareth Exp $

\subsection{Testing the template bank}
\label{sub:testingbank}

\begin{figure}
\begin{center}
\includegraphics[width=0.23\textwidth]{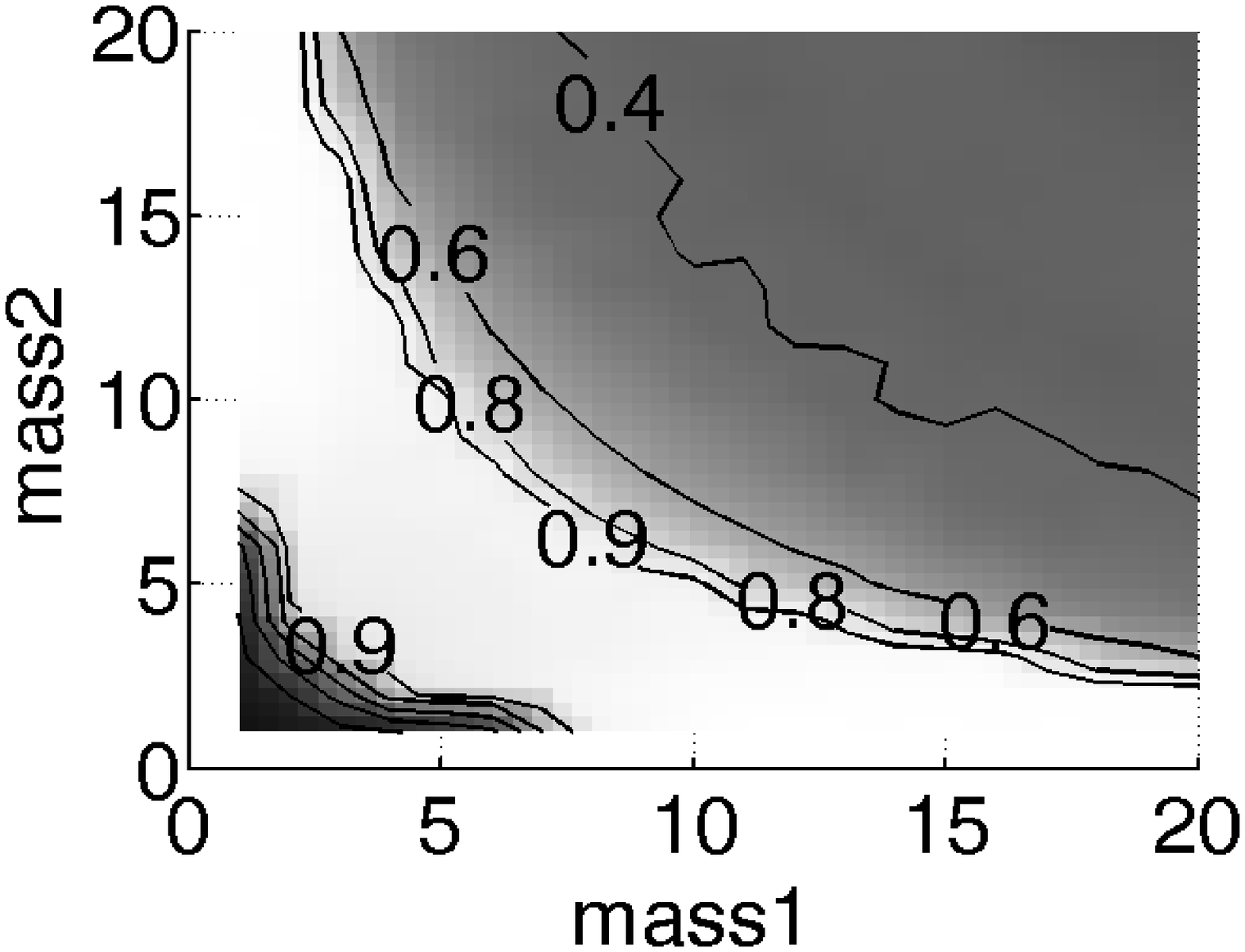}
\includegraphics[width=0.23\textwidth]{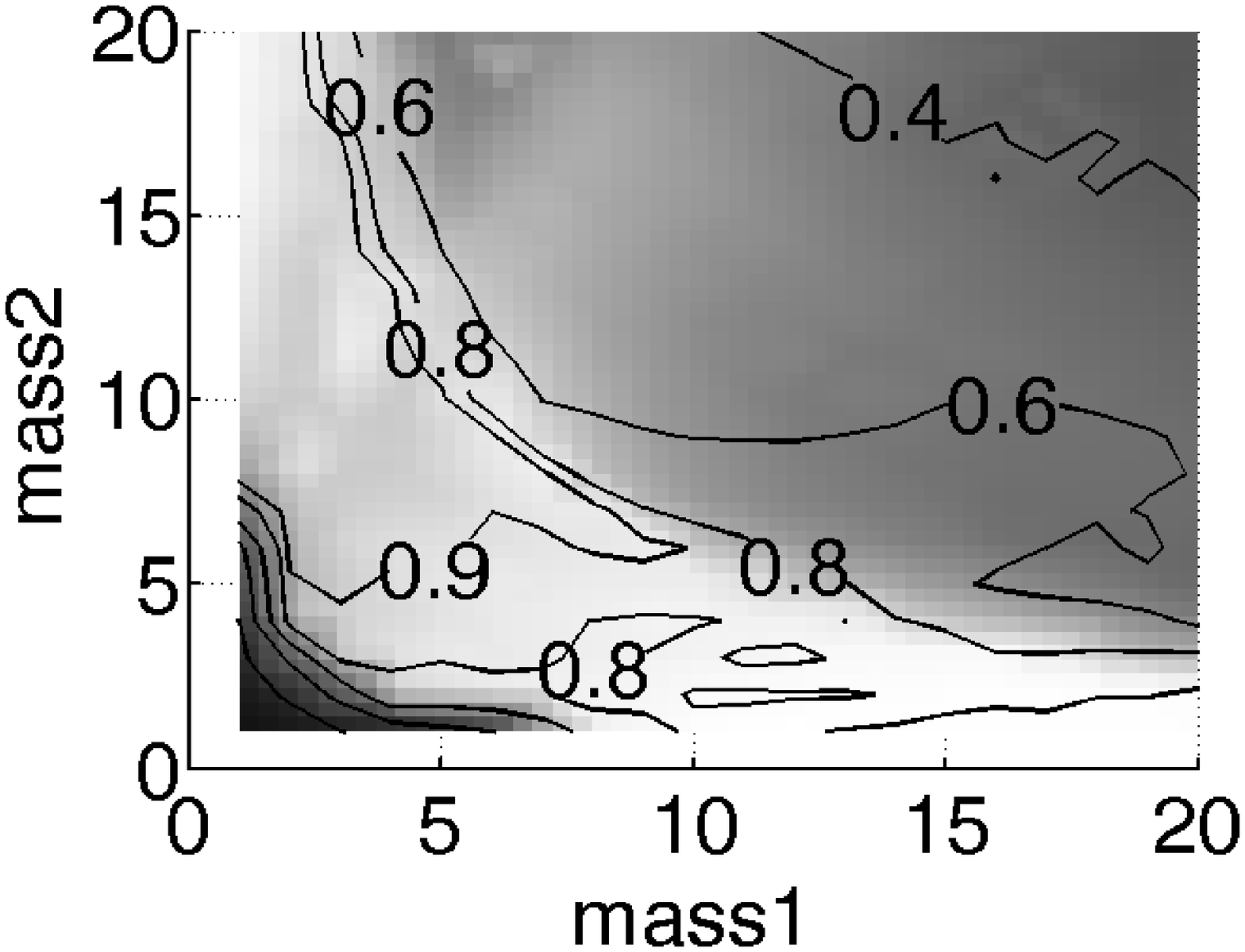} \\
\includegraphics[width=0.23\textwidth]{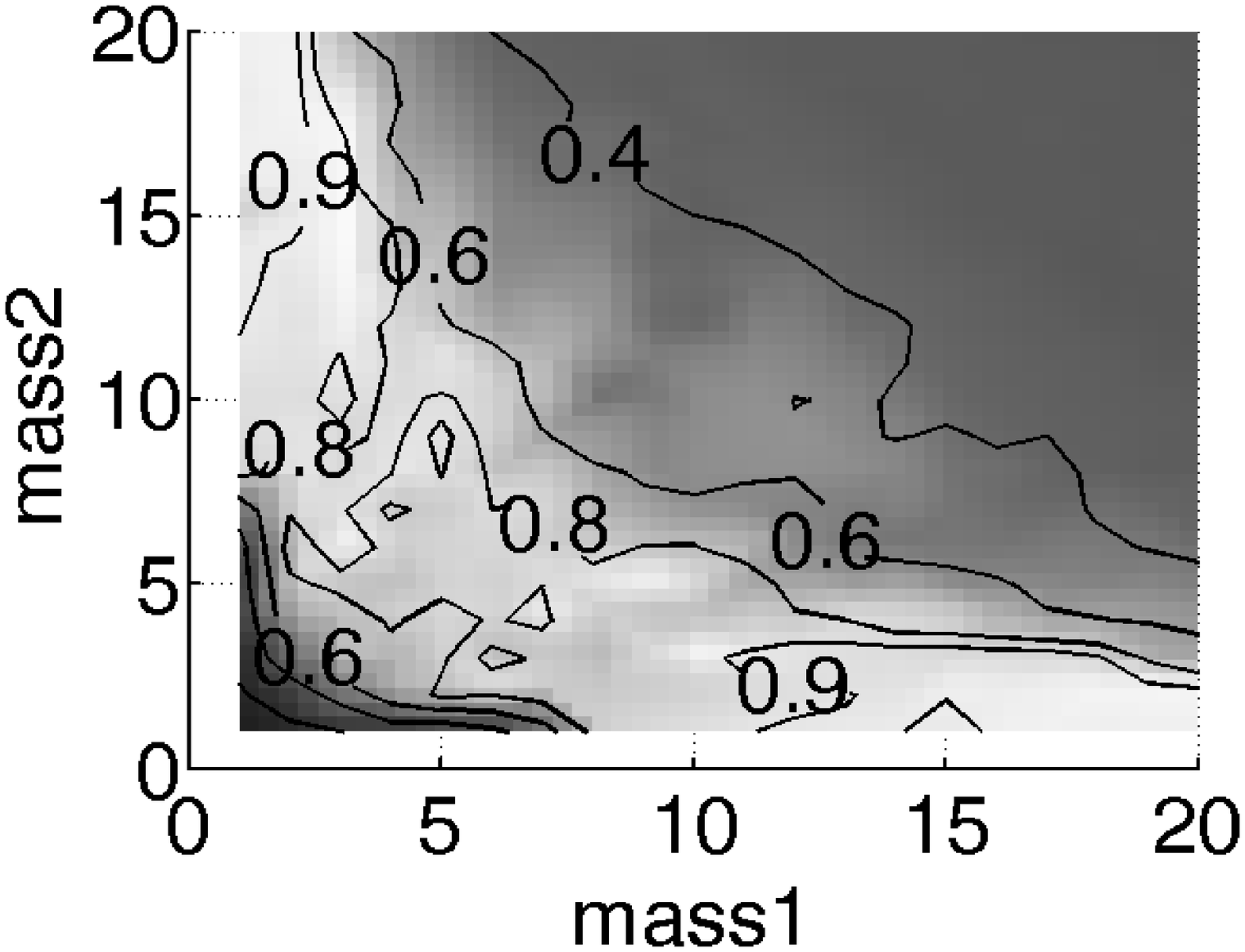} 
\includegraphics[width=0.23\textwidth]{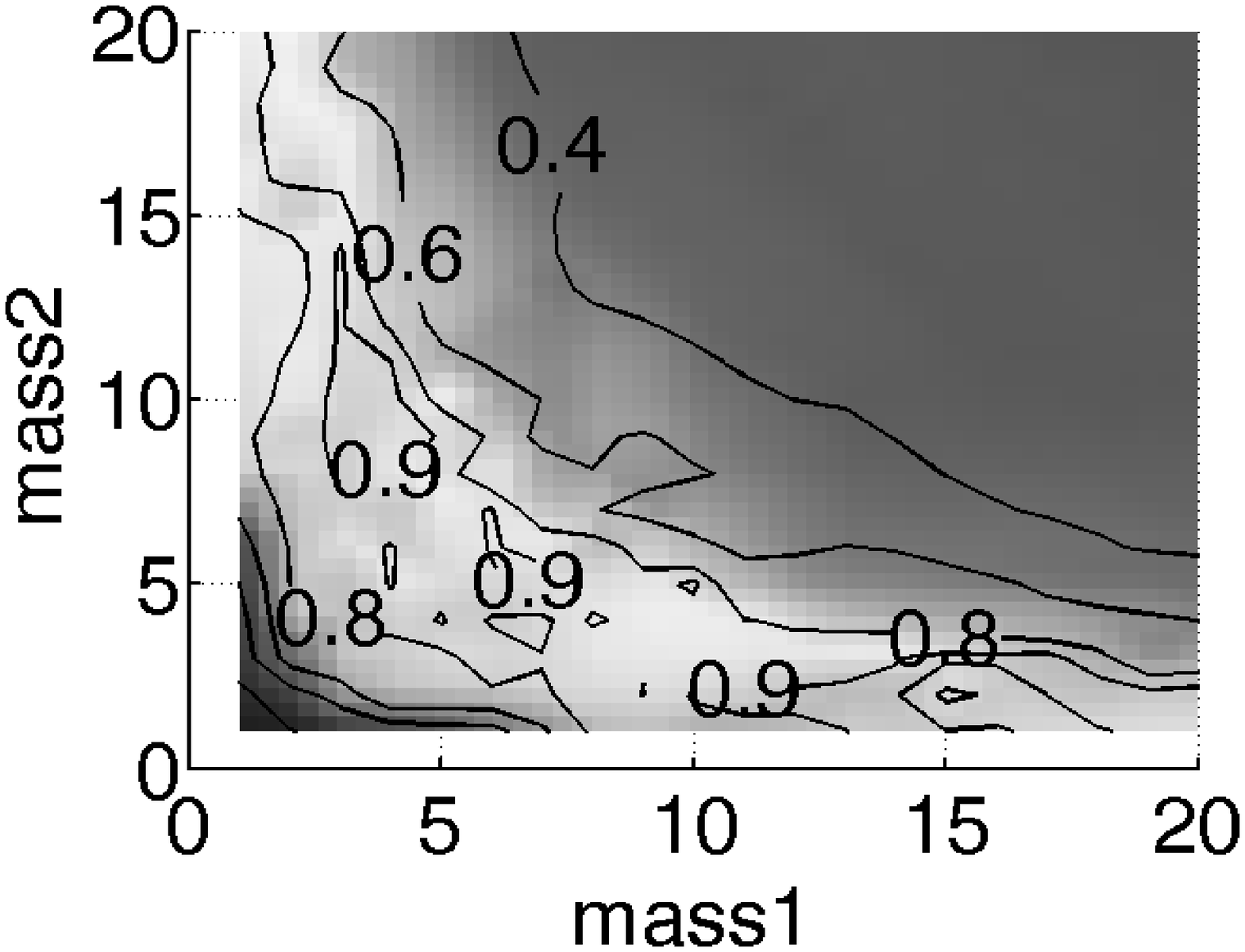}
\end{center}
\caption{Plots showing the best match achieved by filtering a series of simulated signals
through the template bank described in this section.
The values on the x and y axes correspond to the component masses of the binary source to which the simulated
signal corresponds.
The shade of grey in the plots shows the best match achieved for a given simulated signal.
The four subplots correspond to four different spin-configurations of the binary source. 
The top-left subplot shows results for a non-spinning binary system.  
The top-right subplot shows results for a system consisting of one non-spinning object and one maximally
spinning object with its spin slightly misaligned with the orbital angular momentum. We would expect this
system to precess.
The bottow two subplots show results for two generic precessing systems consisting of two maximally 
spinning bodies with spins and orbital angular momentum all misaligned from each other.
We see that the region of the mass plane for which we obtain matches $> 0.9$ is largest for the
non-spinning system and tends to be concentrated in the asymmetric mass region loosely bounded by
$1.0~M_{\odot} < m_{1} < 3.0~M_{\odot}$ and
$12.0~M_{\odot} < m_{2} < 20.0~M_{\odot}$.}
\label{banksims}
\end{figure}

The template bank was tested using a series of simulated signals
constructed using the equations of the target waveforms 
described in Sec.~\ref{sub:target}.
We considered a variety of spin configurations including systems
where neither, one or both bodies were spinning.
We also considered masses outside the range we expected the template
bank to have good coverage in order to fully evaluate the range of masses
for which it could be used.
For each spin configuration we created a series of signals corresponding
to every mass combination: 
%$[1,2...20]\times[1,2...20]M_{\odot}$.
$1.0~M_{\odot} < m_{1}, m_{2} < 20.0~M_{\odot}$.
Using the initial LIGO design sensitivity we then measured the best
match that could be obtained for every signal using our template bank.
Figure~\ref{banksims} shows a sample of the results from the tests of the template
bank.
As expected we found that our template bank achieved the highest matches
for non-spinning (and therefore non-precessing) binaries. 
Performance degrades as spin-precessional effects become more pronounced 
i.e., when both bodies are spinning maximally with spins misaligned from 
the orbital angular momenta.
The template achieved matches $>0.9$ for a mass range 
%$[1-3]\times[12-20] M_{\odot}$.
$1.0~M_{\odot} < m_{1} < 3.0~M_{\odot}$ and
$12.0~M_{\odot} < m_{2} < 20.0~M_{\odot}$ 
(and equivalent systems with $m_{1}$ and $m_{2}$ swapped).
The detection template family (described in Sec.~\ref{sub:dtf}) is capable of obtaining 
high matches for comparable mass systems, the lower matches obtained for comparable mass 
systems are a result of targeting our template bank on asymmetric mass ratio systems 
(which are more susceptible to spin effects and conform to the strong modulation approximation). 

Matches below the specified minimal match of $0.95$ in the bank's region of good coverage
are a consequence of (small) differences between the DTF and the target waveforms
meaning that the DTF {\it cannot} perfectly match the target waveforms.
The {\it fitting factor} (FF) measures the reduction of SNR due to differences between the DTF
and the target waveform \cite{Apostolatos:1995} (and should not be confused with the minimal match
which measures the loss of SNR due to discreteness of the template bank \cite{Owen:1995tm}).  
The DTF performance is evaluated and its fitting factor is measured 
in Sec.~VI of Ref.~\cite{BuonannoChenVallisneri:2003b}, for NS-BH systems an average FF of $\approx 0.93$
was measured \footnote{The authors of Ref.~\cite{BuonannoChenVallisneri:2003b} use a downhill simplex method
called {\tt AMOEBA}\cite{NumericalRecipesInC} in order to obtain the best possible matches between the DTF and the target waveforms. 
This method works well for signals with high SNR but would not be effective in searching for weak signals
in real detector data.}.

 % Gareth
\rcsid$Id: pipeline.tex,v 1.30 2008/06/27 11:04:36 gareth Exp $

\section{Search Pipeline}
\label{sub:pipeline}
The pipeline used for this search is the same as used in the other S3 searches
for binary inspirals \cite{LIGOS3S4all} and is described fully in a set of 
companion papers \cite{findchirppaper,LIGOS3S4Tuning}. 
This pipeline has been significantly updated since the S2 analysis and a
brief summary is now given.

In Sec.~\ref{sub:data} we discuss the S3 data set.
In Sec.~\ref{sub:coincAnalysis} we describe how we decide whether triggers
measured in different detectors could be associated with the same gravitational
wave event.
In Sec.~\ref{sub:combSNR} we introduce the %{\it bitten-lsq} 
statistic which we use to assign SNRs to the events found in coincidence between two or more
detectors.
In Sec.~\ref{sub:background} we describe how we estimate the expected rate
of accidental coincidences.

\subsection{Data sample}
\label{sub:data}
%The Laser Interferometric Gravitational-wave Observatory (LIGO) consist of three
%detectors located at two sites across the US. The LIGO Hanford Observatory
%(LHO) in Washington state consists of two co-located interferometers of
%arm length 4km and 2km and are known as H1 and H2 respectively.
%The LIGO Livingston Observatory (LLO) in Louisiana consists of a single 4km
%interferometer known as L1.
%All three detectors were operated throughout S3 which spanned 70 days (1680 hours)
%between October 31, 2003 and January 9, 2004.
%
To begin with we construct a list of times for which two or more of the detectors
are operating nominally, in what is referred to as {\it science mode}.
By demanding that a gravitational wave be detected in coincidence between two or
more detectors we simultaneously decrease the probability of inferring a detection
when no true signal was present (a false alarm) and improve the confidence we
have in a detection of a true signal. Data collected by the LHO detectors was only
analyzed when both detectors were in science mode. This was due to concerns that
since both of these detectors share the same vacuum system, the laser beam of a
detector in anything but science mode might interfere with the other detector.
%
%The analysis presented here uses data collected only when both of the LHO
%detectors were in science mode.
%In total 778.2 hours of science data collected by the LHO detectors when both
%were in science mode were analysed and 289.4 hours of science data collected by L1
%was analysed.
%The total duration of science data collected when all three detectors were in
%science mode totalled 184.2 hours.
%This left 604.1 hours of science data taken from H1 and H2 when L1 data was not used.

%Throughout this paper we will make use of some terminology that we will introduce
%now. 
We denote periods of time when all three detectors are in science
mode as H1-H2-L1 times and periods when only the Hanford detectors are on as
H1-H2 times. A coincident trigger consisting of a trigger in the H1 detector and the
L1 detector will be referred to as an H1-L1 coincident trigger and similarly for
other combinations of detectors. 
%Therefore an ``H1-L1 event candidate in H1-H2-L1 times''
%refers to an event candidate consisting of an H1 trigger in coincidence with an
%L1 trigger measured during a stretch of time when all three detectors were in
%science mode.

In this search we analyze 
%184.2 
184 hours of H1-H2-L1 data and 
%604.1 
604 hours of H1-H2 data (see Table \ref{tab:analysedtimes}).
During these times we construct template banks for each detector and subsequently
produce a list of triggers whose SNR exceeded our threshold.
%To minimize the false alarm probability we demand that a gravitational wave be observed 
%to have similar parameters (e.g. component masses, time of arrival) 
%in 2 or more detectors.

Around 9\% of the data is specified as {\it playground data} and is used to tune the
various parameters (e.g., SNR thresholds and coincidence windows) used in the full search.
Playground data is not included in the
upper limit calculation but is still searched for possible detections.
We also construct lists of {\it veto times} during which the data we analyze had poor data quality
due to short stretches of instrumental or environmental noise \cite{LIGOS3S4Tuning, Vetoes}. 
All coincident data is analyzed but gravitational wave candidates found during 
veto times will be subjected to greater scrutiny than those found during other times.

%\begin{table}[htdp]
%\caption{Summary of the amount of data analysed in our various data sets. In S3 we
%only analyse data from the LHO detectors when both H1 and H2 are in science mode.
%Around 9\% of the data is classified as {\it playground data} and is used to tune
%the parameters of the search. Playground data is not included in the
%upper limit calculation.  }
%\begin{ruledtabular}
%\begin{tabular}{c|cc}
%Data type & Total analysed (s) & Non-playground (s) \\
%\hline
%H1-H2    & 2174541 & 1971614  \\ 
%H1-H2-L1 & 663129  & 601905  \\ 
%\end{tabular}
%\end{ruledtabular}
%\label{tab:analysedtimes}
%\end{table}

\begin{table}[htdp]
\caption{Summary of the amount of data analyzed in our various data sets. In S3 we
only analyze data from the LHO detectors when both H1 and H2 are in science mode. 
Around 9\% of the data is classified as {\it playground data} and is used to tune
the parameters of the search.}
\begin{ruledtabular}
\begin{tabular}{c|cc}
Data type & Total analyzed (hours) & Non-playground (hours) \\
\hline
H1-H2    & 604   & 548  \\
%H1-H2    & 604.0   & 547.7  \\
H1-H2-L1 & 184   & 167  \\
%H1-H2-L1 & 184.2   & 167.2  \\
\end{tabular}
\end{ruledtabular}
\label{tab:analysedtimes}
\end{table}

\begin{figure}[hbtp]
\includegraphics[width=0.5\textwidth]{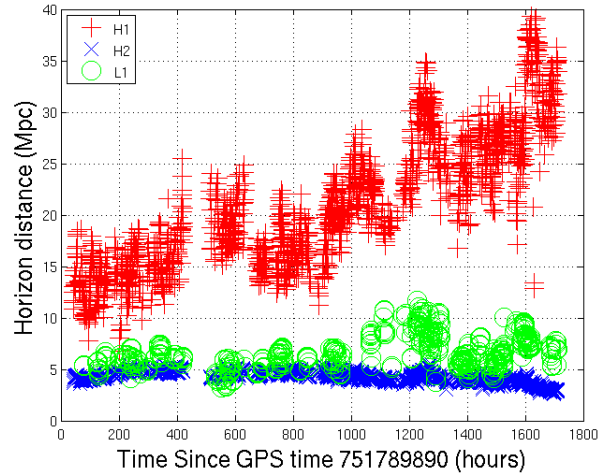}
\caption{Distance to which an optimally oriented non-spinning 
%$(4,10)M_{\odot}$ 
$(2,16)M_{\odot}$ 
binary can be
detected with SNR $= 8$ throughout S3. 
For systems with spinning components, the horizon distance would be equal or less than what is shown in this figure
since any spin-induced precession would cause the system to become less than optimally oriented and therefore
reduce the measured amplitude of its emission.
We see a large improvement in the sensitivity
of H1 during this science run.}
\label{fig:horizondist}
\end{figure}

We can compare the sensitivities of the LIGO detectors by measuring the 
{\it horizon distance} of a particular source --- this is the distance to
which an optimally oriented source can be observed with SNR $= 8$. 
In Fig.~\ref{fig:horizondist} we plot the horizon distance 
of a 
%$(4,10)M_{\odot}$ 
$(2,16)M_{\odot}$ 
binary. This choice of component mass reflects that the
template bank used for this search (see Section \ref{sub:templatebank}) achieves
highest matches for asymmetric binaries.
In Fig.~1 of \cite{LIGOS3S4all} the horizon distance for a range of symmetric
binaries is shown. 
%The improvement in sensitivity of the LIGO detectors
%is indicated by the increase in the horizon distance obtained for a given
%source by each successive science run.

\subsection{Coincident analysis}
\label{sub:coincAnalysis}
To minimize the false alarm probability we demand that a gravitational wave
signal be observed by two or more detectors with similar parameters.
In order to determine whether a trigger measured by one particular detector should
be considered as coincident with a trigger in another detector we define
a set of coincidence windows. 
In this search we demand that for triggers from different detectors to be considered
as coincident they must satisfy the following conditions: 
\begin{eqnarray}
|t_{1} - t_{2}| & < & \Delta t_{1} + \Delta t_{2} + T_{1,2}, \\
|\psi_{0,1} - \psi_{0,2}| & < & \Delta \psi_{0,1} + \Delta \psi_{0,2} \and \\
|\psi_{3,1} - \psi_{3,2}| & < & \Delta \psi_{3,1} + \Delta \psi_{3,2}
\end{eqnarray}
where $t_{i}$, $\psi_{0,i}$ and $\psi_{3,i}$ are the measured time of 
coalescence and phenomenological mass parameters measured using our 
template bank in detector $i$; $\Delta t_{i}$, $\Delta \psi_{0,i}$ and $\Delta \psi_{3,i}$
are our coincidence windows in detector $i$ and $T_{i,j}$ is the light travel time between
detector locations $i$ and $j$. The light travel time between LHO and LLO is $\sim 10$ ms.

We tune our coincidence windows on the playground data in order to 
recover as many of our simulated signals
as possible whilst trying to minimize the false alarm rate. 
The use of playground data allows us to tune our search parameters
without biasing the results of our full analysis.
The tuning method used for this and the non-spinning search on S3/S4
data is described fully in \cite{LIGOS3S4Tuning}.
Using this tuning method we find our coincidence windows to be equal for each 
detector with values $\Delta t = 100$ ms, 
$\Delta \psi_{0} = 40,000~ \rm{Hz}^{5/3} $  and 
$\Delta \psi_{3} = 600~ \rm{Hz}^{2/3}$.
The value of $\Delta t$ used in this search is four times larger than the 
$25$ ms value used in the S3 search for non-spinning binary black holes 
\cite{LIGOS3S4all} indicating that the estimation of arrival time of a 
gravitational waveform is less well determined in this search than in the
non-spinning search. 
%This difference is because the non-spinning search uses multiple (three) values
%of the $f_{\rm{cut}}$ parameter for each point in the $\psi_{0}, \psi_{3}$ space
%whilst this search, in order to reduce number of templates to a reasonable 
%level, uses only a single value of $f_{\rm{cut}}$ for each $\psi_{0}, \psi_{3}$
%combination.
%Since $f_{\rm{cut}}$ determines at what point we should terminate the generation of
%a template it directly affects the measurement of arrival time of a measured
%gravitational wave signal. Fewer values of $f_{\rm{cut}}$ will lead to worse
%estimation of arrival time and therefore require a larger time coincidence
%window $\Delta t$ if simulated signals are to be found in coincidence.

\subsection{Combined SNR}
\label{sub:combSNR}
For coincident triggers we use a combined signal-to-noise ratio $\rho_{c}$
statistic based upon the individual signal-to-noise ratios $\rho_{i}$
measured by each detector:
\begin{equation}
\label{combSNR_2ifo}
%\rho_{c}^{2} = \min \{ \rho_{1}^2 + \rho_{2}^{2} + \rho_{3}^{2},
%a_{1} \rho_{1}^{2} -b_{1},
%a_{2} \rho_{2}^{2} -b_{2},
%a_{3} \rho_{3}^{2} -b_{3}
%\}
\rho_{c}^{2} = {\rm min} \left\{ \sum_i \rho_{\rm i}^2,
(a\rho_{\rm i}-b)^2 \right\}.
\end{equation}
In practice the parameters $a$ and $b$ are tuned so that the contours
of false alarm generated using Eq.~(\ref{combSNR_2ifo}) separate triggers generated
by software injection of simulated signals and background triggers
as cleanly as possible \cite{LIGOS3S4Tuning} (see the next subsection for details
of how we estimate the background). In this search we used values 
$a = b = 3$ for all detectors. For coincident triggers found in all three 
detectors we use:
\begin{equation}
\label{combSNR_3ifo}
\rho_{c}^{2} =  \sum_i \rho_{\rm i}^2.
\end{equation}

\subsection{Background Estimation}
\label{sub:background}
We estimate the rate of accidental coincidences, otherwise known as the
background or false alarm rate, for this search through analysis of 
time-shifted data. We time-shift the triggers obtained from each detector
relative to each other and then repeat our analysis, searching for triggers
that occur in coincidence between 2 or more of the detectors. By choosing
our time-shifts to be suitably large 
($\gg 10$ ms light travel time between LHO and LLO)  
we ensure that none of the coincident triggers identified
in our time-shift analysis could be caused by a true gravitational wave signal
and can therefore be used as an estimate of the rate of accidental coincidences.
In practice we leave H1 data unshifted and time-shift H2 and L1 by increments
of $10$ and $5$ s respectively.
In this search, we analyzed 100 sets of time-shifted data 
(50 forward shifts and 50 backward shifts). 
For clarity we will use the term {\it in-time} to mean triggers which have not
been time-shifted.
 % Gareth
\rcsid$Id: vetoes.tex,v 1.15 2007/12/07 17:58:38 gareth Exp $
\section{Vetoes}
\label{sub:vetoes}

\subsection{Instrument based vetoes}
We are able to veto some background triggers by observing correlation between the
gravitational wave channel (AS\_Q) of a particular detector and one or more of its auxiliary channels 
which monitor the local physical environment. Since we would not expect a true gravitational wave signal
to excite the auxiliary channels, we will treat as suspicious any excitation in the gravitational wave
channel that is coincident in time with excitations in the auxiliary channels.
A list of auxiliary channels found to effectively veto spurious (non-gravitational wave coincident triggers) 
were identified and used for all S3 searches \cite{Vetoes}.
%Additionally, another two 
%auxiliary channels, L1:LSC-REFL\_Q and H1:LSC-MICH\_CTRL channels were investigated as potential vetoes 
%for this search. 
%However the total amount of data discounted when using these channels as 
%vetoes, known as the {\it dead-time}, was unacceptably large 
%and it was decided not to use them in this search. 
%Consequently, only the standard list of auxiliary channels were used to veto
%coincident triggers in this search. 
Additional vetoes based upon other auxiliary channels were considered but were subsequently abandoned because the total
amount of data these channels would have discounted, known as the {\it dead-time}, was unacceptably large.

\subsection{Signal based vetoes}
We can use the fact that the Hanford detectors are co-located to veto coincident triggers
whose measured amplitude is not consistent between H1 and H2.
%We do not have an effective distance calculation for the triggers 
%arising from this search but instead look 
We check for consistency between the SNR values
measured using H1 and H2 data for triggers found in coincidence. Since H1 is 
%generally
the more sensitive instrument we simply required that the SNR measured in H1 be greater
than that measured in H2 for an event to survive this veto.
Since data from H1 and H2 was only analysed when both were in science mode, this veto means that
there will be no H2-L1 coincident triggers since this would indicate that H2 had detected a trigger
which H1 was unable to detect.

The $\chi^{2}$ veto used for the primordial black hole and binary neutron star 
searches~\cite{LIGOS3S4all} has not not been investigated for use in searches using
detection template families (i.e., this search and the S2-S4 searches for 
non-spinning binary black holes \cite{LIGOS2bbh, LIGOS3S4all}).  

%The $\chi^{2}$ veto is not used in either of the black hole searches
%(spinning or non-spinning) for two reasons: i) for systems consisting
%of larger masses the emitted gravitational wave will be short reducing
%validity of a $\chi^{2}$ test and perhaps more importantly ii)
%we do not know the effects of using the $\chi^2$ veto with detection template
%families. It is known however that even very small mismatches between
%template and signal can lead to large $\chi^2$ values.

\rcsid$Id: results.tex,v 1.27 2008/04/03 10:34:17 gareth Exp $
\section{Results}
\label{sub:results}

In the search of the S3 LIGO data described in this paper, 
%no coincident triggers were found in all
%three detectors. 
no triple-coincident event candidates 
(exceeding our pre-determined SNR threshold and satisfying the
coincidence requirements described in Sec.~\ref{sub:coincAnalysis})
were found in triple-time (H1-H2-L1) data.
Many double-coincident event candidates were found in both triple-time
and double-time (H1-H2) data.

%\begin{table*}[htdp]
%\caption{Characteristics of the loudest coincident event found in the
%entire S3 data set, irrespective of the coincident times. This loudest
%event is used in the final upper limit calculations. 
%The first column is the combined snr found using the bitten lsq statistic.
%The second column is the fraction of time shifts with a
%louder candidate than the one found in the non time-shifted data. 
%The remaining columns give the parameters that were recorded in each detector.
%{\bf include gps? remove non-physical params?}}
%\begin{ruledtabular}
%\begin{tabular}{cc|cccc|cccc}
%&& & H1 &&& & H2 && \\
%$\rho_{c}$ 
%& $1-P_{\rm B}(\rho_{c})$ 
%& $\rho$
%& $\psi_{0}$ 
%& $\psi_{3}$ 
%& $\beta$  
%& $\rho$
%& $\psi_{0}$ 
%& $\psi_{3}$ 
%& $\beta$ \\
%\hline
%58.297
%& 0.23 
%& 119.302 
%& 384304.688 
%& -3937.092 
%& 51.827
%& 20.432  
%& 399101.906
%& -4086.411 
%& 560.839 
%\end{tabular}
%\end{ruledtabular}
%\label{tab:Candidates}
%\end{table*}
%\subsection{Comparison between in-time triggers with estimated background}
%Figure \ref{slide_trigs} shows the number of (double-coincident) triggers found in each time-shift
%of the H1-H2 data set.
%We find that the number of H1-H2 double-coincident triggers found during
%in-time H1-H2 times is consistent with the number of coincident triggers yielded by
%the time-shift analysis. Similarly we find that the number
%of in-time H1-H2 and H1-L1 coincident triggers found when all three detectors were in
%science mode is also consistent with the numbers predicted by the time-shift analysis.

\begin{figure}[hbtp]
\includegraphics[width=0.5\textwidth]{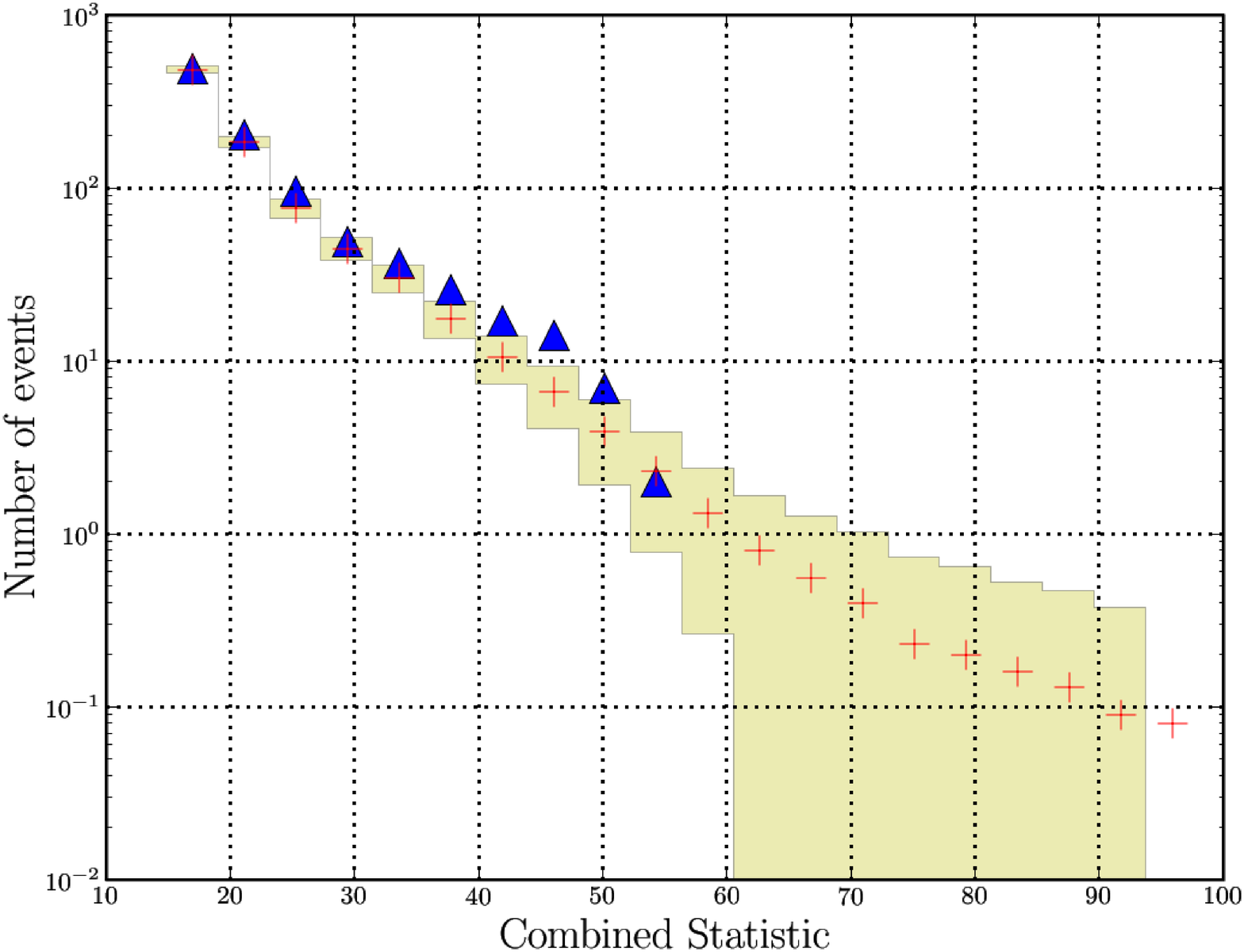}
\caption{
Cumulative histograms of the combined SNR, $\rho_{c}$ for in-time coincident triggers
(triangles) and our background (crosses with one-sigma deviation shown)
for all H1-H2 and H1-H2-L1 times within S3.
We see a small excess in the number of in-time coincident triggers with combined SNR
$\sim 45$. This excess was investigated and was caused by an excess of H1-H2 coincident
triggers. Since H1 and H2 are co-located, both detectors are affected by the same
local disturbances (e.g., seismic activity) which contributes to the number of in-time
coincidences but which is under-represented in time-shift estimates of the background.
}
\label{cum_hist}
\end{figure}

A cumulative histogram of combined SNR for in-time and background coincident triggers is shown in
Fig.~\ref{cum_hist}. 
We see that, at the SNR threshold 
(i.e., the leftmost points on this figure), 
the number of in-time double-coincident triggers is consistent with the number
of coincident triggers yielded by the time-shift analysis.
The small excess in the number of in-time H1-H2 coincident triggers at higher SNRs indicates
that there is some correlation between the LHO detectors. The coincident triggers contributing
to this excess have been investigated and are not believed to be
caused by gravitational waves. 
Seismic activity at the Hanford site has been recorded throughout S3 and 
can cause data to become noisy simultaneously in H1 and H2.  
Coincident triggers caused by seismic noise will predominantly
cause only in-time coincidences 
(although time-shift coincidences caused by two seismic 
events separated in time but shifted together can occur)
leading to an excess of in-time coincident triggers as we have observed in Fig.~\ref{cum_hist}.
As mentioned previously, there were no coincident triggers observed by all three 
detectors.

\begin{figure}[hbtp]
\includegraphics[width=0.5\textwidth]{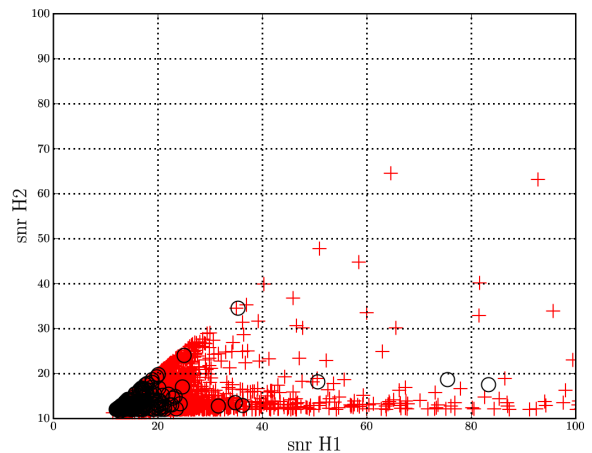}
\caption{Scatter plot of SNR for coincident triggers in H1-H2 times. 
The black circles represent in-time coincident triggers and the 
light colored (red) pluses represent 
time-shift coincident triggers that we use to estimate the background. 
Note that due to our signal based veto on H1/H2 SNR
we see no coincident triggers with $\rho_{\mathrm{H1}} < \rho_{\mathrm{H2}}$.}
\label{snr_snr}
\end{figure}

A scatter plot of the SNRs measured for coincident triggers in H1-H2 times
is shown in Fig.~\ref{snr_snr}. The distribution of our in-time triggers 
is consistent with our estimation of the background. This is also true for
the double-coincident triggers measured in H1-H2-L1 times.

The loudest in-time coincident trigger was observed in H1-H2 when
only the Hanford detectors were in science mode.
This event candidate is measured to have SNRs of
$119.3$ in H1, $20.4$ in H2 and a combined SNR of $58.3$.
The loudest coincident triggers are subjected to systematic follow-up investigations
in which a variety of information (e.g., data quality at time of triggers, 
correlation between the detector's auxiliary channels and the gravitational wave channel)
is used to assess whether the coincident triggers could be confidently claimed as
detection of gravitational wave events.
This event is found at a time flagged for ``conditional''
vetoing. This means that during these times some of the detectors
auxiliary channels exhibited correlation with the
gravitational wave channel (AS\_Q ) and that we should be careful
in how we treat event candidates found in these times.
For this particular coincident trigger an auxiliary channel indicated an
increased numbers of dust particles passing through the dark port beam
of the interferometer~\cite{Vetoes}.
Upon further investigation it was found that this coincident trigger occurred
during a period of seismic activity at the Hanford site and we
subsequently discounted this candidate as a potential gravitational wave event.
Time-frequency images of the gravitational wave channel around the time of this
candidate were inconsistent with expectations of what an inspiral signal should look like,
further reducing the plausibility of this candidate being a true gravitational wave event.
It is interesting, but unsurprising, to note that during the search for non-spinning
binary black holes that also used S3 LIGO data, high-SNR triggers associated with this seismic activity were also detected \cite{LIGOS3S4all}.
Furthermore, the 20 next loudest event candidates were also investigated and none 
were found to be plausible gravitational wave event candidates.
Work is in progress to automate the follow-up investigative procedure and to include
new techniques including null-stream and Markov chain Monte Carlo analysis for assessing
the plausibility of coincident triggers as gravitational wave events. 

 % Gareth
\rcsid$Id: upperlimits.tex,v 1.28 2008/04/03 10:34:17 gareth Exp $
\section{Upper Limits}
\label{sub:upperlimit}

\begin{figure}[hbtp]
%\begin{center}
\includegraphics[width=0.5\textwidth]{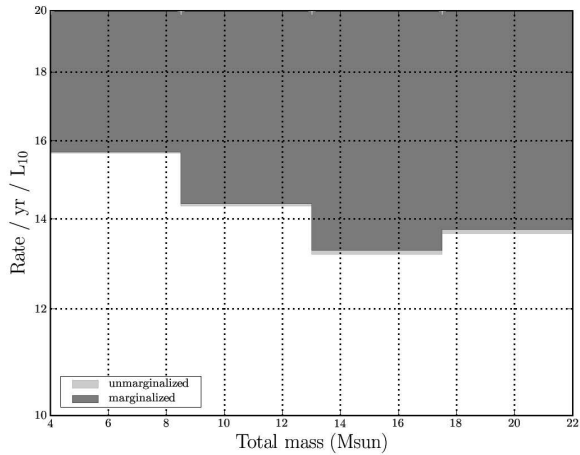}
%\end{center}
\caption{Upper limits on the spinning binary coalescence rate per
$\mathrm{L}_{10}$ as a function of the total mass of the binary. 
For this calculation, we have evaluated the efficiency of the search 
using a population of binary systems with $m_1 = 1.35 M_{\odot}$
and $m_2$ uniformly distributed between $2$ and $20 M_{\odot}$.
The darker area on the plot
shows the region excluded after marginalization over the estimated systematic errors
whereas the lighter region shows the region excluded if these systematic errors are 
ignored. The effect of marginalization is typically small ($<1\%$).
The initial decrease in the upper limit corresponds to the
increasing amplitude of the signals as total mass increases. 
The subsequent increase in upper limit is due to the counter effect that as 
total mass increases the signals become shorter and have fewer cycles in LIGO's 
frequency band of good sensitivity.}
\label{fig:upperlimit}
\end{figure}

Given the absence of plausible detection candidates within the search
described above, we have calculated an upper limit on the rate of
spinning compact object coalescence in the universe. 
We quote the upper limit rate in units of $\rm{yr}^{-1} \rm{L}_{10}^{-1}$
where $\rm{L}_{10} = 10^{10} \, \rm{L}_{\odot,B}$ is $10^{10}$ times the blue 
light luminosity of the sun.

The absorption-corrected blue light luminosity of a galaxy infers its massive star formation rate which
we assume scales with the rate of compact binary coalescence within it \cite{Phinney:1991ei}. 
This assumption is well justified when the galaxies reached by the detector are dominated
by spiral galaxies with ongoing star formation (e.g., the Milky Way).
Results papers reporting on S1 and S2 ~\cite{LIGOS1iul, LIGOS2iul, LIGOS2bbh} 
have quoted the upper limit in units of Milky Way Equivalent Galaxy (MWEG) which 
is equivalent to about $1.7 \, \rm{L}_{10}$.
Upper limits on the rate of coalescences calculated during other searches
using S3 and S4 LIGO are given in units of $\rm{L}_{10}$ \cite{LIGOS3S4all}.

The upper limit calculations are based on the loudest event 
statistic~\cite{loudestGWDAW03,ul}, which uses both the detection 
efficiency at the combined SNR of the loudest event candidate and the 
associated background probability.
The in-time non-playground data set (which we use to set the upper limit) is
{\it blinded} in the sense that all analysis parameters are tuned 
(as described in Secs.~\ref{sub:pipeline}) prior to its analysis.

The Bayesian upper limit at a confidence level $\alpha$, assuming a uniform
prior on the rate $R$, is given by \cite{ul}
\begin{equation}
  1 - \alpha = e^{-R\, T \, \mathcal{C}_{L}(\rho_{c,\rm{max}})}  
    \left[ 1 + 
    \left( \frac{\Lambda}{1+\Lambda} \right)
    \, R\, T \, \mathcal{C}_{L}(\rho_{c,\rm{max}}) \right] \\
    \label{bayesianprob}
\end{equation}
where $\mathcal{C}_{L}(\rho_{c,\rm{max}})$ is the cumulative blue light luminosity to which 
we are sensitive at a given value of combined SNR $\rho_{c,\rm{max}}$, $T$ is the
observation time, and $\Lambda$ is a measure of the likelihood that the 
loudest event is consistent with being a signal and
inconsistent with background (as estimated using time-shifts).
We evaluate the cumulative luminosity $\mathcal{C}_{L}$ at the combined SNR
of the loudest coincident trigger seen in this search, $\rho_{c,\rm{max}} = 58.3$ 
(see Sec.~\ref{sub:results} for discussion of this coincident trigger). 
The expression for $\Lambda$ is
\begin{equation}\label{xi}
  \Lambda =  
  \frac{|\mathcal{C}_{L}^{\prime}(\rho_{c,\rm{max}})|}{P_{B}^{\prime}(\rho_{c,\rm{max}})}
  \left[
  \frac{\mathcal{C}_{L}(\rho_{c,\rm{max}})}{P_{B}(\rho_{c,\rm{max}})}
  \right]^{-1}
  \, ,
\end{equation}
where the derivatives are with respect to $\rho_{c}$.  
$P_{\rm B}(\rho)$ is the probability that
all background coincident triggers (as estimated using time-shifts) have a combined SNR
less than $\rho$. 
For the loudest event candidate in this search we find $P_{\rm B} = 0.23$
and $\Lambda = 0.05$.
In the case where the loudest event candidate is most likely due to the 
background $\Lambda \rightarrow 0$ and the upper limit becomes
\begin{equation}\label{ul_bg}
  R_{90\%} = \frac{2.3}{ T \, \mathcal{C}_{L}(\rho_{c,\rm{max}})} \, .
\end{equation}
In the limit of zero background, i.e., the event is definitely not background, 
$\Lambda \rightarrow \infty$ and the numerator in Eq.~(\ref{ul_bg}) becomes
$3.9$. The observation time $T$ is taken from Table~\ref{tab:analysedtimes},
where we use the analyzed time {\em not} in the playground.
This is consistent with our blind analysis strategy. 

In searches for systems consisting of non-spinning bodies efficiency is typically 
found as a function of its effective distance and chirp mass \cite{systematics}. 
For a system consisting of non-spinning bodies effective distance can be calculated
using the distance to the source, its inclination with respect to the
detector and the detector's antenna response functions (see Eq.~(2) of \cite{LIGOS3S4all} and
\cite{thorne.k:1987}).
For a system consisting of spinning bodies, its inclination with respect to a detector
will evolve during the course of the inspiral making the calculation of effective
distance complicated.
Instead, in this search we find efficiency and predicted source luminosity as a
function of the inverse of the {\it expected SNR} of a source.
The expected SNR is defined as the SNR that would be obtained for a given simulated
source assuming we use a template that perfectly matches the emitted gravitational waveform
and a detector whose noise power spectrum we can estimate accurately. 
By taking the inverse of the expected SNR we obtain a quantity which behaves similarly
to the effective distance by taking larger values for signals which are nearer and/or optimally
oriented to the detector and thus more easily detectable and by taking smaller values as 
the signals become less detectable.

Following the tests of the template bank (Sec.~\ref{sub:testingbank}) we also know that
the efficiency at which we are able to detect sources will depend on their spins as well 
as their effective distance and component masses.
In this upper limit calculation we assess the efficiency of the search using software injection
of simulated signals representing a population of sources with spins randomized so that 
i) the spin magnitude of each of the compact objects is distributed uniformly in the 
range $0 < \chi < 1$ and 
ii) the direction of compact object's spin is uniformly distributed on the surface 
of a sphere. 
%The masses of the sources used to assess the efficiency are in the 
%range $1.0~M_{\odot} < m_{1}, m_{2} < 20.0~M_{\odot}$.
The distances of the simulated sources are chosen uniformly on a logarithmic
scale.
The sky-positions and initial polarization and inclination angles of the
simulated sources are all chosen randomly and to be uniformly
distributed on the surface of a sphere.
We evaluated the efficiency of this search for masses in the range 
$1.0~M_{\odot} < m_{1}, m_{2} < 20.0~M_{\odot}$. 
During S3, LIGO's efficiency was dominated by sources within the 
Milky Way for which detection efficiency was high 
across the entire mass range investigated due to the proximity of 
these sources to Earth. 
We also had some detection efficiency for binaries in M31 and M33.

The cumulative luminosity $\mathcal{C}_{L}(\rho_{c})$ can be obtained by generating a population 
of simulated signals using information on the observed distribution of sources from standard 
astronomy catalogs. 
We use a model based on~\cite{LIGOS3S4Galaxies} for the distribution
of blue light luminosity throughout the nearby Universe. 
We use software injection of simulated signals (the target waveforms
described in Sec.~\ref{sub:target}) to evaluate the efficiency $\mathcal{E}$
for observing an event with combined SNR greater than $\rho_{c}$, as a
function of the source's expected SNR.
We then integrate $\mathcal{E}$ times the predicted
source luminosity $L$ as a function of expected SNR and mass.  
Since a binary system will generally have slightly different orientations with 
respect to the two LIGO observatory sites, the detectors at the two sites will 
both measure slightly different expected SNRs.
The source's luminosity and the efficiency with which it is detected are functions of 
both expected SNRs, and the integration needed is two-dimensional:
\begin{widetext}
\begin{equation}
  \mathcal{C}_{L}(\rho_{c})=
  \int_0^{\infty} 
  \int_0^{\infty} 
  \mathcal{E}(D_{\rho,\rm{H}},D_{\rho,\rm{L}},\rho) \, L(D_{\rho,\rm{H}}, dD_{\rho,\rm{L}}) \,
  dD_{\rho,\rm{L}} \, dD_{\rho,\rm{H}} 
\end{equation}
\end{widetext}
where $D_{\rho}$ is the distance measure equal to the inverse of the expected SNR, 
at LHO (H) or LLO (L). As mentioned earlier, we evaluate $\mathcal{C}_{L}$ at
$\rho_{c,\rm{max}} = 58.3$. 
The cumulative luminosity was measured to be $\sim 1.9 L_{10}$ and is dominated
by the Milky Way ($1.7 L_{10}$) with the remainder made up by M31 and M33.

We calculate the upper limit on the rate of coalescence for proto-typical 
NS-BH binaries with masses $m_{1} \sim 1.35~M_{\odot}$ and $m_{2} \sim 5~M_{\odot}$. 
These values correspond to a population of NS-BH binaries with component masses 
similar to those used to assess the NS-NS and BH-BH upper limits in \cite{LIGOS3S4all}. 
To calculate this upper limit we evaluate the efficiency of our search using 
binaries with a Gaussian mass distribution with means, 
$m_{1}= 1.35~M_{\odot}$ and $m_{2} = 5~M_{\odot}$ with standard deviations 
$\sigma_{1}=0.04~M_{\odot}$ and $\sigma_{2} = 1~M_{\odot}$.
%
%The efficiency for this search is computed using a Gaussian mass distribution, 
%with means $m_1 = 1.35 M_{\odot}$, $m_2=5 M_{\odot}$ and standard deviations 
%$\sigma_1 = 0.04 M_{\odot}$, $\sigma_2 = 1 M_{\odot}$.
%These values correspond to a population of NS-BH binaries with component masses similar to 
%those used to assess the NS-NS and BH-BH upper limits in \cite{LIGOS3S4all}.
%%This choice of masses reflects our expectations about
%%the type of binary systems we believe LIGO is most likely to observe based upon
%%population studies~\cite{Belczynski:2002}.
%%The choice of relatively low component masses also means that our upper limit
%%estimate should be conservative.  
These efficiencies are measured with simulated injected signals, using the 
same pipeline we used to find our candidates, counting the number of injections 
detected with SNR above $\rho_{c,\rm{max}}$, and the number missed.  
Assuming a Gaussian distribution of masses, we obtain an upper
limit of 
$\mathcal{R}_{90\%} =
%12.68 
15.8
\,\mathrm{yr}^{-1}\,\mathrm{L_{10}}^{-1}$. 
The upper limit calculation takes into account the possible systematic
uncertainties which 
arise in this search, which are described in some detail in~\cite{systematics}, 
and we will follow the analysis presented there to calculate the
systematic errors for the above result.  The most significant effects are due to
the possible calibration inaccuracies of the detectors (estimated 
using hardware injections of simulated signals) and the finite number of Monte Carlo injections
performed. 

We must also evaluate the systematic errors associated with the
chosen astrophysical model of potential sources within the galaxy.  We obtain
upper limits on the rate after marginalization over the estimated systematic
errors, as described in~\cite{systematics,ul}. 
After marginalization over these errors
we obtain an upper limit of 
$\mathcal{R}_{90\%} =
15.9 
\,\mathrm{yr}^{-1}\,\mathrm{L_{10}}^{-1}$.
%Marginalization over these errors causes only a very small (less than $0.5\%$) 
%increase to the upper limit.
 
%We also calculated the upper limits with uniform total mass distribution. 
We also calculate upper limits for a range of binary systems with $m_1 = 1.35 M_{\odot}$
and $m_2$ uniformly distributed between $2$ and $20 M_{\odot}$.
These upper limits, both before and after marginalization are shown 
in Fig.~\ref{fig:upperlimit}. 
%It can be seen from this figure that the upper limit decreases as the total mass 
%of the binaries increases. This effect is partially due to improved template bank coverage
%for heavier asymmetric systems (highest matches for generic spinning systems are obtained for binaries
%with masses $m_1 \simeq 2 M_{\odot}$ and $m_2 \simeq 16 M_{\odot}$, see Sec.~\ref{sub:testingbank}) but
%is primarily caused by the increase of SNR with total mass 
%($\rho \propto M^{5/6} \eta^{1/2} / d$, see e.g., \cite{thorne.k:1987}).   
%Since systems with {\it higher} SNR are more likely to be detected we are able to place {\it lower} upper limits
%on the rate of coalescence of these systems in the case where no detection was made.
%During S3 the region of the Universe from which we would be able to detect binary inspirals
%was dominated by the Milky Way. 
%Consequently we find that the detection efficiency measured for this search is almost uniform across 
%the range of component masses investigated (i.e., $1.0~M_{\odot} < m_{1}, m_{2} < 20.0~M_{\odot}$).  
%The upper limits calculated are almost constant across the range of masses plotted with a slight 
%increase in the upper limits for binaries with larger total mass where the signals become shorter.
 
 % Gareth
\rcsid$Id: conclusions.tex,v 1.16 2008/06/23 15:30:13 gareth Exp $
\section{Conclusions}
\label{sub:conclusions}

In this paper we have described the first search for gravitational waves emitted during the inspiral of 
compact binaries with spinning component bodies, which was carried out using data taken during the third LIGO science run.
Interaction between the binary's orbital angular momentum and the spin angular momenta of its components
will cause precession of its orbital plane resulting in the modulation of the observed gravitational wave.

This search uses a detection template family designed specially to capture the spin-induced modulations of
the gravitational waveform which could have resulted in them being missed by other searches
targeted at non-spinning systems. The search pipeline used to carry out this and the other
recent inspiral searches has been significantly improved since S2 and is fully described in a 
companion paper~\cite{LIGOS3S4all}.

There were no plausible gravitational wave event candidates detected within the 788 hours of
S3 data analyzed. The upper limit on the rate of coalescence for prototypical NS-BH binaries with spinning component 
bodies was calculated to be 
$\mathcal{R}_{90\%} =
15.9 \,\mathrm{yr}^{-1}\,\mathrm{L_{10}}^{-1}$ 
once errors had been marginalized over.

%Preliminary work is underway on an improved search for binaries with 
%spinning component bodies that will use S5 data
%which is greater in sensitivity and observation time than previous data sets.
%
The S5 LIGO data is greatly improved in sensitivity and observation time
than previous data sets. 
There is considerable work in progress to further improve the techniques used to
search for binaries with spinning component bodies using the S5 data.
This includes development of an improved parameter-space metric which does not 
depend on the strong modulation approximation and allows us to 
search a larger region of parameter space, including binaries with comparable 
masses. 
Preparation for another search for binaries with spinning component bodies 
using a template family described by physical (rather than phenomenological) 
parameters \cite{Pan:2003qt} is also underway.
 % Gareth and Alessandra
%-- Appendix: filtering details

\acknowledgments

% new acknowledgment from http://www.lsc-group.phys.uwm.edu/ppcomm/

The authors gratefully acknowledge the support of the United States
National Science Foundation for the construction and operation of the
LIGO Laboratory and the Science and Technology Facilities Council of the
United Kingdom, the Max-Planck-Society, and the State of
Niedersachsen/Germany for support of the construction and operation of
the GEO600 detector. The authors also gratefully acknowledge the support
of the research by these agencies and by the Australian Research Council,
the Council of Scientific and Industrial Research of India, the Istituto
Nazionale di Fisica Nucleare of Italy, the Spanish Ministerio de
Educaci\'on y Ciencia, the Conselleria d'Economia, Hisenda i Innovaci\'o of
the Govern de les Illes Balears, the Scottish Funding Council, the
Scottish Universities Physics Alliance, The National Aeronautics and
Space Administration, the Carnegie Trust, the Leverhulme Trust, the David
and Lucile Packard Foundation, the Research Corporation, and the Alfred
P. Sloan Foundation.
This paper was assigned LIGO document number LIGO-P070102-06-Z.

\bibliography{paper}

\begin{thebibliography}{59}
\expandafter\ifx\csname natexlab\endcsname\relax\def\natexlab#1{#1}\fi
\expandafter\ifx\csname bibnamefont\endcsname\relax
  \def\bibnamefont#1{#1}\fi
\expandafter\ifx\csname bibfnamefont\endcsname\relax
  \def\bibfnamefont#1{#1}\fi
\expandafter\ifx\csname citenamefont\endcsname\relax
  \def\citenamefont#1{#1}\fi
\expandafter\ifx\csname url\endcsname\relax
  \def\url#1{\texttt{#1}}\fi
\expandafter\ifx\csname urlprefix\endcsname\relax\def\urlprefix{URL }\fi
\providecommand{\bibinfo}[2]{#2}
\providecommand{\eprint}[2][]{\url{#2}}

\bibitem[{\citenamefont{Abbott et~al.}(2004{\natexlab{a}})}]{LIGOS1instpaper}
\bibinfo{author}{\bibfnamefont{B.}~\bibnamefont{Abbott}} \bibnamefont{et~al.}
  (\bibinfo{collaboration}{{LIGO} Scientific Collaboration}),
  \bibinfo{journal}{Nucl. Instrum. Methods} \textbf{\bibinfo{volume}{A517}},
  \bibinfo{pages}{154} (\bibinfo{year}{2004}{\natexlab{a}}).

\bibitem[{\citenamefont{Barish and Weiss}(1999)}]{Barish:1999}
\bibinfo{author}{\bibfnamefont{B.~C.} \bibnamefont{Barish}} \bibnamefont{and}
  \bibinfo{author}{\bibfnamefont{R.}~\bibnamefont{Weiss}},
  \bibinfo{journal}{Phys.\ Today} \textbf{\bibinfo{volume}{52 (Oct)}},
  \bibinfo{pages}{44} (\bibinfo{year}{1999}).

\bibitem[{\citenamefont{L{\"u}ck and the {GEO}600~team}(1997)}]{Luck:1997hv}
\bibinfo{author}{\bibfnamefont{H.}~\bibnamefont{L{\"u}ck}} \bibnamefont{and}
  \bibinfo{author}{\bibnamefont{the {GEO}600~team}}, \bibinfo{journal}{Class.
  Quant. Grav.} \textbf{\bibinfo{volume}{14}}, \bibinfo{pages}{1471}
  (\bibinfo{year}{1997}), \bibinfo{note}{\textit{ibid}, \textbf{23}, S71
  (2006)}.

\bibitem[{\citenamefont{Acernese et~al.}(2006)\citenamefont{Acernese, Amico,
  Alshourbagy, Antonucci, Aoudia, Avino, Babusci, Ballardin, Barone, Barsotti
  et~al.}}]{0264-9381-23-19-S01}
\bibinfo{author}{\bibfnamefont{F.}~\bibnamefont{Acernese}},
  \bibinfo{author}{\bibfnamefont{P.}~\bibnamefont{Amico}},
  \bibinfo{author}{\bibfnamefont{M.}~\bibnamefont{Alshourbagy}},
  \bibinfo{author}{\bibfnamefont{F.}~\bibnamefont{Antonucci}},
  \bibinfo{author}{\bibfnamefont{S.}~\bibnamefont{Aoudia}},
  \bibinfo{author}{\bibfnamefont{S.}~\bibnamefont{Avino}},
  \bibinfo{author}{\bibfnamefont{D.}~\bibnamefont{Babusci}},
  \bibinfo{author}{\bibfnamefont{G.}~\bibnamefont{Ballardin}},
  \bibinfo{author}{\bibfnamefont{F.}~\bibnamefont{Barone}},
  \bibinfo{author}{\bibfnamefont{L.}~\bibnamefont{Barsotti}},
  \bibnamefont{et~al.}, \bibinfo{journal}{Classical and Quantum Gravity}
  \textbf{\bibinfo{volume}{23}}, \bibinfo{pages}{S635} (\bibinfo{year}{2006}),
  \urlprefix\url{http://stacks.iop.org/0264-9381/23/S635}.

\bibitem[{\citenamefont{Thorne}(1987)}]{thorne.k:1987}
\bibinfo{author}{\bibfnamefont{K.~S.} \bibnamefont{Thorne}}, in
  \emph{\bibinfo{booktitle}{Three hundred years of gravitation}}, edited by
  \bibinfo{editor}{\bibfnamefont{S.~W.} \bibnamefont{Hawking}}
  \bibnamefont{and} \bibinfo{editor}{\bibfnamefont{W.}~\bibnamefont{Israel}}
  (\bibinfo{publisher}{Cambridge University Press},
  \bibinfo{address}{Cambridge}, \bibinfo{year}{1987}),
  chap.~\bibinfo{chapter}{9}, pp. \bibinfo{pages}{330--458}.

\bibitem[{\citenamefont{Grishchuk et~al.}(2001)\citenamefont{Grishchuk,
  Lipunov, Postnov, Prokhorov, and Sathyaprakash}}]{grishchuk-2001-171}
\bibinfo{author}{\bibfnamefont{L.~P.} \bibnamefont{Grishchuk}},
  \bibinfo{author}{\bibfnamefont{V.~M.} \bibnamefont{Lipunov}},
  \bibinfo{author}{\bibfnamefont{K.~A.} \bibnamefont{Postnov}},
  \bibinfo{author}{\bibfnamefont{M.~E.} \bibnamefont{Prokhorov}},
  \bibnamefont{and} \bibinfo{author}{\bibfnamefont{B.~S.}
  \bibnamefont{Sathyaprakash}}, \bibinfo{journal}{USP.FIZ.NAUK}
  \textbf{\bibinfo{volume}{171}}, \bibinfo{pages}{3} (\bibinfo{year}{2001}),
  \urlprefix\url{http://arxiv.org/abs/astro-ph/0008481}.

\bibitem[{\citenamefont{Abbott et~al.}(2007)}]{LIGOS3S4all}
\bibinfo{author}{\bibfnamefont{B.}~\bibnamefont{Abbott}} \bibnamefont{et~al.}
  (\bibinfo{collaboration}{{LIGO} Scientific Collaboration}),
  \bibinfo{journal}{gr-qc}  (\bibinfo{year}{2007}), \eprint{arXiv:0704.3368, to
  be submitted to Phys.~Rev.~D}.

\bibitem[{\citenamefont{{Portegies Zwart} and
  McMillan}(2000)}]{PortegiesZwart:2000}
\bibinfo{author}{\bibfnamefont{S.~F.} \bibnamefont{{Portegies Zwart}}}
  \bibnamefont{and} \bibinfo{author}{\bibfnamefont{S.~L.~W.}
  \bibnamefont{McMillan}}, \bibinfo{journal}{Astrophys. J.}
  \textbf{\bibinfo{volume}{528}}, \bibinfo{pages}{L17} (\bibinfo{year}{2000}).

\bibitem[{\citenamefont{{O'Leary} et~al.}(2006)\citenamefont{{O'Leary},
  {Rasio}, {Fregeau}, {Ivanova}, and {O'Shaughnessy}}}]{2006ApJForS3S4Joint}
\bibinfo{author}{\bibfnamefont{R.~M.} \bibnamefont{{O'Leary}}},
  \bibinfo{author}{\bibfnamefont{F.~A.} \bibnamefont{{Rasio}}},
  \bibinfo{author}{\bibfnamefont{J.~M.} \bibnamefont{{Fregeau}}},
  \bibinfo{author}{\bibfnamefont{N.}~\bibnamefont{{Ivanova}}},
  \bibnamefont{and}
  \bibinfo{author}{\bibfnamefont{R.}~\bibnamefont{{O'Shaughnessy}}},
  \bibinfo{journal}{Astrophysical Journal} \textbf{\bibinfo{volume}{637}},
  \bibinfo{pages}{937} (\bibinfo{year}{2006}), \eprint{astro-ph/0508224. See
  also Kim et al., arXiv:0608280, and Kalogera et al., arXiv:0612144.}

\bibitem[{\citenamefont{{Fregeau} et~al.}(2006)\citenamefont{{Fregeau},
  {Larson}, {Miller}, {O'Shaughnessy}, and {Rasio}}}]{imbhlisa-2006}
\bibinfo{author}{\bibfnamefont{J.~M.} \bibnamefont{{Fregeau}}},
  \bibinfo{author}{\bibfnamefont{S.~L.} \bibnamefont{{Larson}}},
  \bibinfo{author}{\bibfnamefont{M.~C.} \bibnamefont{{Miller}}},
  \bibinfo{author}{\bibfnamefont{R.}~\bibnamefont{{O'Shaughnessy}}},
  \bibnamefont{and} \bibinfo{author}{\bibfnamefont{F.~A.}
  \bibnamefont{{Rasio}}}, \bibinfo{journal}{Astrophysical Journal Letters}
  \textbf{\bibinfo{volume}{646}}, \bibinfo{pages}{L135} (\bibinfo{year}{2006}),
  \eprint{astro-ph/0605732}.

\bibitem[{\citenamefont{{O'Shaughnessy}
  et~al.}(2005{\natexlab{a}})\citenamefont{{O'Shaughnessy}, {Kim}, {Fragos},
  {Kalogera}, and {Belczynski}}}]{OShaughnessy:2005}
\bibinfo{author}{\bibfnamefont{R.}~\bibnamefont{{O'Shaughnessy}}},
  \bibinfo{author}{\bibfnamefont{C.}~\bibnamefont{{Kim}}},
  \bibinfo{author}{\bibfnamefont{T.}~\bibnamefont{{Fragos}}},
  \bibinfo{author}{\bibfnamefont{V.}~\bibnamefont{{Kalogera}}},
  \bibnamefont{and}
  \bibinfo{author}{\bibfnamefont{K.}~\bibnamefont{{Belczynski}}},
  \bibinfo{journal}{\apj} \textbf{\bibinfo{volume}{633}}, \bibinfo{pages}{1076}
  (\bibinfo{year}{2005}{\natexlab{a}}), \eprint{arXiv:astro-ph/0504479}.

\bibitem[{\citenamefont{O'Shaughnessy et~al.}(2006)\citenamefont{O'Shaughnessy,
  Kim, Kalogera, and Belczynski}}]{OShaughnessy:2006b}
\bibinfo{author}{\bibfnamefont{R.}~\bibnamefont{O'Shaughnessy}},
  \bibinfo{author}{\bibfnamefont{C.}~\bibnamefont{Kim}},
  \bibinfo{author}{\bibfnamefont{V.}~\bibnamefont{Kalogera}}, \bibnamefont{and}
  \bibinfo{author}{\bibfnamefont{K.}~\bibnamefont{Belczynski}}
  (\bibinfo{year}{2006}), \eprint{astro-ph/0610076}.

\bibitem[{\citenamefont{Abbott et~al.}(2006)}]{LIGOS2bbh}
\bibinfo{author}{\bibfnamefont{B.}~\bibnamefont{Abbott}} \bibnamefont{et~al.}
  (\bibinfo{collaboration}{{LIGO} Scientific Collaboration}),
  \bibinfo{journal}{Phys. Rev. D} \textbf{\bibinfo{volume}{73}},
  \bibinfo{pages}{062001} (\bibinfo{year}{2006}), \eprint{gr-qc/0509129}.

\bibitem[{\citenamefont{Helmstrom}(1968)}]{helmstrom-1968}
\bibinfo{author}{\bibfnamefont{C.~W.} \bibnamefont{Helmstrom}},
  \emph{\bibinfo{title}{Statistical Theory of Signal Detection, 2nd edition}}
  (\bibinfo{publisher}{Pergamon Press, London}, \bibinfo{year}{1968}).

\bibitem[{\citenamefont{Apostolatos et~al.}(1994)\citenamefont{Apostolatos,
  Cutler, Sussman, and Thorne}}]{Apostolatos:1994}
\bibinfo{author}{\bibfnamefont{T.~A.} \bibnamefont{Apostolatos}},
  \bibinfo{author}{\bibfnamefont{C.}~\bibnamefont{Cutler}},
  \bibinfo{author}{\bibfnamefont{G.~J.} \bibnamefont{Sussman}},
  \bibnamefont{and} \bibinfo{author}{\bibfnamefont{K.~S.}
  \bibnamefont{Thorne}}, \bibinfo{journal}{Phys. Rev. D}
  \textbf{\bibinfo{volume}{49}}, \bibinfo{pages}{6274} (\bibinfo{year}{1994}).

\bibitem[{\citenamefont{Kidder}(1995)}]{kidder:821}
\bibinfo{author}{\bibfnamefont{L.~E.} \bibnamefont{Kidder}},
  \bibinfo{journal}{Physical Review D (Particles, Fields, Gravitation, and
  Cosmology)} \textbf{\bibinfo{volume}{52}}, \bibinfo{pages}{821}
  (\bibinfo{year}{1995}),
  \urlprefix\url{http://link.aps.org/abstract/PRD/v52/p821}.

\bibitem[{\citenamefont{Apostolatos}(1995)}]{Apostolatos:1995}
\bibinfo{author}{\bibfnamefont{T.~A.} \bibnamefont{Apostolatos}},
  \bibinfo{journal}{Phys. Rev. D} \textbf{\bibinfo{volume}{52}},
  \bibinfo{pages}{605} (\bibinfo{year}{1995}).

\bibitem[{\citenamefont{Apostolatos}(1996)}]{Apostolatos:1996rg}
\bibinfo{author}{\bibfnamefont{T.~A.} \bibnamefont{Apostolatos}},
  \bibinfo{journal}{Phys. Rev. D} \textbf{\bibinfo{volume}{54}},
  \bibinfo{pages}{2438} (\bibinfo{year}{1996}).

\bibitem[{\citenamefont{Buonanno
  et~al.}(2003{\natexlab{a}})\citenamefont{Buonanno, Chen, and
  Vallisneri}}]{BuonannoChenVallisneri:2003b}
\bibinfo{author}{\bibfnamefont{A.}~\bibnamefont{Buonanno}},
  \bibinfo{author}{\bibfnamefont{Y.}~\bibnamefont{Chen}}, \bibnamefont{and}
  \bibinfo{author}{\bibfnamefont{M.}~\bibnamefont{Vallisneri}},
  \bibinfo{journal}{Phys. Rev. D} \textbf{\bibinfo{volume}{67}},
  \bibinfo{pages}{104025} (\bibinfo{year}{2003}{\natexlab{a}}),
  \bibinfo{note}{erratum-ibid. 74 (2006) 029904(E)}.

\bibitem[{\citenamefont{Grandcl\'{e}ment
  et~al.}(2003)\citenamefont{Grandcl\'{e}ment, Kalogera, and
  Vecchio}}]{GrandclementKalogeraVecchio:2003}
\bibinfo{author}{\bibfnamefont{P.}~\bibnamefont{Grandcl\'{e}ment}},
  \bibinfo{author}{\bibfnamefont{V.}~\bibnamefont{Kalogera}}, \bibnamefont{and}
  \bibinfo{author}{\bibfnamefont{A.}~\bibnamefont{Vecchio}},
  \bibinfo{journal}{Phys. Rev. D} \textbf{\bibinfo{volume}{67}},
  \bibinfo{pages}{042003} (\bibinfo{year}{2003}).

\bibitem[{\citenamefont{Barker and O'Connell}(1975)}]{BarkerOConnell1975}
\bibinfo{author}{\bibfnamefont{B.~M.} \bibnamefont{Barker}} \bibnamefont{and}
  \bibinfo{author}{\bibfnamefont{R.~F.} \bibnamefont{O'Connell}},
  \bibinfo{journal}{Phys. Rev. D} \textbf{\bibinfo{volume}{12}},
  \bibinfo{pages}{329} (\bibinfo{year}{1975}).

\bibitem[{\citenamefont{Belczynski et~al.}(2002)\citenamefont{Belczynski,
  Kalogera, and Bulik}}]{Belczynski:2002}
\bibinfo{author}{\bibfnamefont{K.}~\bibnamefont{Belczynski}},
  \bibinfo{author}{\bibfnamefont{V.}~\bibnamefont{Kalogera}}, \bibnamefont{and}
  \bibinfo{author}{\bibfnamefont{T.}~\bibnamefont{Bulik}},
  \bibinfo{journal}{Astrophys. J.} \textbf{\bibinfo{volume}{572}},
  \bibinfo{pages}{407} (\bibinfo{year}{2002}).

\bibitem[{\citenamefont{Kalogera}(2000)}]{Kalogera:2000}
\bibinfo{author}{\bibfnamefont{V.}~\bibnamefont{Kalogera}},
  \bibinfo{journal}{Astrophys. J.} \textbf{\bibinfo{volume}{541}},
  \bibinfo{pages}{042003} (\bibinfo{year}{2000}).

\bibitem[{\citenamefont{{O'Shaughnessy}
  et~al.}(2005{\natexlab{b}})\citenamefont{{O'Shaughnessy}, {Kaplan},
  {Kalogera}, and {Belczynski}}}]{2005ApJ...632.1035O}
\bibinfo{author}{\bibfnamefont{R.}~\bibnamefont{{O'Shaughnessy}}},
  \bibinfo{author}{\bibfnamefont{J.}~\bibnamefont{{Kaplan}}},
  \bibinfo{author}{\bibfnamefont{V.}~\bibnamefont{{Kalogera}}},
  \bibnamefont{and}
  \bibinfo{author}{\bibfnamefont{K.}~\bibnamefont{{Belczynski}}},
  \bibinfo{journal}{Astrophysical Journal} \textbf{\bibinfo{volume}{632}},
  \bibinfo{pages}{1035} (\bibinfo{year}{2005}{\natexlab{b}}),
  \eprint{astro-ph/0503219}.

\bibitem[{\citenamefont{Babak et~al.}(2006)\citenamefont{Babak,
  Balasubramanian, Churches, Cokelaer, and Sathyaprakash}}]{BBCCS:2006}
\bibinfo{author}{\bibfnamefont{S.}~\bibnamefont{Babak}},
  \bibinfo{author}{\bibnamefont{Balasubramanian}},
  \bibinfo{author}{\bibfnamefont{D.}~\bibnamefont{Churches}},
  \bibinfo{author}{\bibfnamefont{T.}~\bibnamefont{Cokelaer}}, \bibnamefont{and}
  \bibinfo{author}{\bibfnamefont{B.}~\bibnamefont{Sathyaprakash}},
  \bibinfo{journal}{Class. Quant. Grav.} \textbf{\bibinfo{volume}{23}},
  \bibinfo{pages}{5477} (\bibinfo{year}{2006}), \eprint{gr-qc/0604037}.

\bibitem[{\citenamefont{Allen et~al.}(2005)\citenamefont{Allen, Anderson,
  Brady, Brown, and Creighton}}]{findchirppaper}
\bibinfo{author}{\bibfnamefont{B.~A.} \bibnamefont{Allen}},
  \bibinfo{author}{\bibfnamefont{W.~G.} \bibnamefont{Anderson}},
  \bibinfo{author}{\bibfnamefont{P.~R.} \bibnamefont{Brady}},
  \bibinfo{author}{\bibfnamefont{D.~A.} \bibnamefont{Brown}}, \bibnamefont{and}
  \bibinfo{author}{\bibfnamefont{J.~D.~E.} \bibnamefont{Creighton}}
  (\bibinfo{year}{2005}), \eprint{gr-qc/0509116}.

\bibitem[{\citenamefont{Grandcl\'{e}ment
  et~al.}(2004)\citenamefont{Grandcl\'{e}ment, Ihm, Kalogera, and
  Belczynski}}]{grandclement:102002}
\bibinfo{author}{\bibfnamefont{P.}~\bibnamefont{Grandcl\'{e}ment}},
  \bibinfo{author}{\bibfnamefont{M.}~\bibnamefont{Ihm}},
  \bibinfo{author}{\bibfnamefont{V.}~\bibnamefont{Kalogera}}, \bibnamefont{and}
  \bibinfo{author}{\bibfnamefont{K.}~\bibnamefont{Belczynski}},
  \bibinfo{journal}{Physical Review D (Particles, Fields, Gravitation, and
  Cosmology)} \textbf{\bibinfo{volume}{69}}, \bibinfo{eid}{102002}
  (pages~\bibinfo{numpages}{10}) (\bibinfo{year}{2004}),
  \urlprefix\url{http://link.aps.org/abstract/PRD/v69/e102002}.

\bibitem[{\citenamefont{Thorne}(1974)}]{Thorne:1974ve}
\bibinfo{author}{\bibfnamefont{K.~S.} \bibnamefont{Thorne}},
  \bibinfo{journal}{Astrophys. J.} \textbf{\bibinfo{volume}{191}},
  \bibinfo{pages}{507} (\bibinfo{year}{1974}).

\bibitem[{\citenamefont{Cook et~al.}(1994)\citenamefont{Cook, Shapiro, and
  Teukolsky}}]{Cook:1993qr}
\bibinfo{author}{\bibfnamefont{G.~B.} \bibnamefont{Cook}},
  \bibinfo{author}{\bibfnamefont{S.~L.} \bibnamefont{Shapiro}},
  \bibnamefont{and} \bibinfo{author}{\bibfnamefont{S.~A.}
  \bibnamefont{Teukolsky}}, \bibinfo{journal}{Astrophys. J.}
  \textbf{\bibinfo{volume}{424}}, \bibinfo{pages}{823} (\bibinfo{year}{1994}).

\bibitem[{\citenamefont{Remillard and McClintock}(2006)}]{remillard-2006-44}
\bibinfo{author}{\bibfnamefont{R.~A.} \bibnamefont{Remillard}}
  \bibnamefont{and} \bibinfo{author}{\bibfnamefont{J.~E.}
  \bibnamefont{McClintock}}, \bibinfo{journal}{Annual Review of Astronomy and
  Astrophysics} \textbf{\bibinfo{volume}{44}}, \bibinfo{pages}{49}
  (\bibinfo{year}{2006}),
  \urlprefix\url{http://www.citebase.org/abstract?id=oai:arXiv.org:astro-ph/06%
06352}.

\bibitem[{\citenamefont{McClintock et~al.}(2006)\citenamefont{McClintock,
  Shafee, Narayan, Remillard, Davis, and Li}}]{mcclintock-2006-652}
\bibinfo{author}{\bibfnamefont{J.~E.} \bibnamefont{McClintock}},
  \bibinfo{author}{\bibfnamefont{R.}~\bibnamefont{Shafee}},
  \bibinfo{author}{\bibfnamefont{R.}~\bibnamefont{Narayan}},
  \bibinfo{author}{\bibfnamefont{R.~A.} \bibnamefont{Remillard}},
  \bibinfo{author}{\bibfnamefont{S.~W.} \bibnamefont{Davis}}, \bibnamefont{and}
  \bibinfo{author}{\bibfnamefont{L.-X.} \bibnamefont{Li}},
  \bibinfo{journal}{The Astrophysical Journal} \textbf{\bibinfo{volume}{652}},
  \bibinfo{pages}{518} (\bibinfo{year}{2006}),
  \urlprefix\url{http://www.citebase.org/abstract?id=oai:arXiv.org:astro-ph/06%
06076}.

\bibitem[{\citenamefont{Clifton and Weisberg}(2008)}]{Clifton-2008}
\bibinfo{author}{\bibfnamefont{T.}~\bibnamefont{Clifton}} \bibnamefont{and}
  \bibinfo{author}{\bibfnamefont{J.~M.} \bibnamefont{Weisberg}},
  \bibinfo{journal}{The Astrophysical Journal} \textbf{\bibinfo{volume}{679}},
  \bibinfo{pages}{687} (\bibinfo{year}{2008}),
  \eprint{http://www.journals.uchicago.edu/doi/pdf/10.1086/587049},
  \urlprefix\url{http://www.journals.uchicago.edu/doi/abs/10.1086/587049}.

\bibitem[{\citenamefont{Blanchet et~al.}(1995)\citenamefont{Blanchet, Damour,
  Iyer, Will, and Wiseman}}]{Blanchet:1995ez}
\bibinfo{author}{\bibfnamefont{L.}~\bibnamefont{Blanchet}},
  \bibinfo{author}{\bibfnamefont{T.}~\bibnamefont{Damour}},
  \bibinfo{author}{\bibfnamefont{B.~R.} \bibnamefont{Iyer}},
  \bibinfo{author}{\bibfnamefont{C.~M.} \bibnamefont{Will}}, \bibnamefont{and}
  \bibinfo{author}{\bibfnamefont{A.~G.} \bibnamefont{Wiseman}},
  \bibinfo{journal}{Phys. Rev. Lett.} \textbf{\bibinfo{volume}{74}},
  \bibinfo{pages}{3515} (\bibinfo{year}{1995}).

\bibitem[{\citenamefont{Blanchet et~al.}(1996)\citenamefont{Blanchet, Iyer,
  Will, and Wiseman}}]{Blanchet:1996pi}
\bibinfo{author}{\bibfnamefont{L.}~\bibnamefont{Blanchet}},
  \bibinfo{author}{\bibfnamefont{B.~R.} \bibnamefont{Iyer}},
  \bibinfo{author}{\bibfnamefont{C.~M.} \bibnamefont{Will}}, \bibnamefont{and}
  \bibinfo{author}{\bibfnamefont{A.~G.} \bibnamefont{Wiseman}},
  \bibinfo{journal}{Class. Quant. Grav.} \textbf{\bibinfo{volume}{13}},
  \bibinfo{pages}{575} (\bibinfo{year}{1996}).

\bibitem[{\citenamefont{Blanchet}(1996)}]{PhysRevD.54.1417}
\bibinfo{author}{\bibfnamefont{L.}~\bibnamefont{Blanchet}},
  \bibinfo{journal}{Phys. Rev. D} \textbf{\bibinfo{volume}{54}},
  \bibinfo{pages}{1417} (\bibinfo{year}{1996}).

\bibitem[{\citenamefont{Blanchet et~al.}(2002)\citenamefont{Blanchet, Faye,
  Iyer, and Joguet}}]{Blanchet:2001ax}
\bibinfo{author}{\bibfnamefont{L.}~\bibnamefont{Blanchet}},
  \bibinfo{author}{\bibfnamefont{G.}~\bibnamefont{Faye}},
  \bibinfo{author}{\bibfnamefont{B.~R.} \bibnamefont{Iyer}}, \bibnamefont{and}
  \bibinfo{author}{\bibfnamefont{B.}~\bibnamefont{Joguet}},
  \bibinfo{journal}{Phys. Rev. D} \textbf{\bibinfo{volume}{65}},
  \bibinfo{pages}{061501(R)} (\bibinfo{year}{2002}), \eprint{gr-qc/0105099}.

\bibitem[{\citenamefont{Kidder et~al.}(1993)\citenamefont{Kidder, Will, and
  Wiseman}}]{PhysRevD.47.R4183}
\bibinfo{author}{\bibfnamefont{L.~E.} \bibnamefont{Kidder}},
  \bibinfo{author}{\bibfnamefont{C.~M.} \bibnamefont{Will}}, \bibnamefont{and}
  \bibinfo{author}{\bibfnamefont{A.~G.} \bibnamefont{Wiseman}},
  \bibinfo{journal}{Phys. Rev. D} \textbf{\bibinfo{volume}{47}},
  \bibinfo{pages}{R4183} (\bibinfo{year}{1993}).

\bibitem[{\citenamefont{Damour et~al.}(2001)\citenamefont{Damour, Iyer, and
  Sathyaprakash}}]{Damour:2000zb}
\bibinfo{author}{\bibfnamefont{T.}~\bibnamefont{Damour}},
  \bibinfo{author}{\bibfnamefont{B.~R.} \bibnamefont{Iyer}}, \bibnamefont{and}
  \bibinfo{author}{\bibfnamefont{B.~S.} \bibnamefont{Sathyaprakash}},
  \bibinfo{journal}{Phys. Rev. D} \textbf{\bibinfo{volume}{63}},
  \bibinfo{pages}{044023} (\bibinfo{year}{2001}).

\bibitem[{\citenamefont{Finn and Chernoff}(1993)}]{FinnChernoff:1993}
\bibinfo{author}{\bibfnamefont{L.}~\bibnamefont{Finn}} \bibnamefont{and}
  \bibinfo{author}{\bibfnamefont{D.}~\bibnamefont{Chernoff}},
  \bibinfo{journal}{Phys. Rev. D} \textbf{\bibinfo{volume}{47}},
  \bibinfo{pages}{2198} (\bibinfo{year}{1993}).

\bibitem[{\citenamefont{Grandcl\'{e}ment and
  Kalogera}(2003)}]{GrandclementKalogera:2003}
\bibinfo{author}{\bibfnamefont{P.}~\bibnamefont{Grandcl\'{e}ment}}
  \bibnamefont{and} \bibinfo{author}{\bibfnamefont{V.}~\bibnamefont{Kalogera}},
  \bibinfo{journal}{Phys. Rev. D} \textbf{\bibinfo{volume}{67}},
  \bibinfo{pages}{082002} (\bibinfo{year}{2003}), \eprint{gr-qc/0211075}.

\bibitem[{\citenamefont{Buonanno et~al.}(2005)\citenamefont{Buonanno, Chen,
  Pan, Tagoshi, and Vallisneri}}]{Buonanno:2005pt}
\bibinfo{author}{\bibfnamefont{A.}~\bibnamefont{Buonanno}},
  \bibinfo{author}{\bibfnamefont{Y.}~\bibnamefont{Chen}},
  \bibinfo{author}{\bibfnamefont{Y.}~\bibnamefont{Pan}},
  \bibinfo{author}{\bibfnamefont{H.}~\bibnamefont{Tagoshi}}, \bibnamefont{and}
  \bibinfo{author}{\bibfnamefont{M.}~\bibnamefont{Vallisneri}},
  \bibinfo{journal}{Phys. Rev. D} \textbf{\bibinfo{volume}{72}},
  \bibinfo{pages}{084027} (\bibinfo{year}{2005}), \eprint{gr-qc/0508064}.

\bibitem[{\citenamefont{Owen}(1996)}]{Owen:1995tm}
\bibinfo{author}{\bibfnamefont{B.~J.} \bibnamefont{Owen}},
  \bibinfo{journal}{Phys. Rev. D} \textbf{\bibinfo{volume}{53}},
  \bibinfo{pages}{6749} (\bibinfo{year}{1996}).

\bibitem[{\citenamefont{Owen and Sathyaprakash}(1999)}]{Owen:1998dk}
\bibinfo{author}{\bibfnamefont{B.~J.} \bibnamefont{Owen}} \bibnamefont{and}
  \bibinfo{author}{\bibfnamefont{B.~S.} \bibnamefont{Sathyaprakash}},
  \bibinfo{journal}{Phys. Rev. D} \textbf{\bibinfo{volume}{60}},
  \bibinfo{pages}{022002} (\bibinfo{year}{1999}).

\bibitem[{\citenamefont{Pan et~al.}(2004)\citenamefont{Pan, Buonanno, Chen, and
  Vallisneri}}]{Pan:2003qt}
\bibinfo{author}{\bibfnamefont{Y.}~\bibnamefont{Pan}},
  \bibinfo{author}{\bibfnamefont{A.}~\bibnamefont{Buonanno}},
  \bibinfo{author}{\bibfnamefont{Y.-b.} \bibnamefont{Chen}}, \bibnamefont{and}
  \bibinfo{author}{\bibfnamefont{M.}~\bibnamefont{Vallisneri}},
  \bibinfo{journal}{Phys. Rev. D} \textbf{\bibinfo{volume}{69}},
  \bibinfo{pages}{104017} (\bibinfo{year}{2004}), \bibinfo{note}{erratum-ibid.
  74 (2006) 029905(E)}, \eprint{gr-qc/0310034}.

\bibitem[{\citenamefont{Abramovici et~al.}(1992)}]{Abramovici:1992ah}
\bibinfo{author}{\bibfnamefont{A.}~\bibnamefont{Abramovici}}
  \bibnamefont{et~al.}, \bibinfo{journal}{Science}
  \textbf{\bibinfo{volume}{256}}, \bibinfo{pages}{325} (\bibinfo{year}{1992}).

\bibitem[{\citenamefont{Arun et~al.}(2005)\citenamefont{Arun, Iyer,
  Sathyaprakash, and Sundararajan}}]{arun:084008}
\bibinfo{author}{\bibfnamefont{K.~G.} \bibnamefont{Arun}},
  \bibinfo{author}{\bibfnamefont{B.~R.} \bibnamefont{Iyer}},
  \bibinfo{author}{\bibfnamefont{B.~S.} \bibnamefont{Sathyaprakash}},
  \bibnamefont{and} \bibinfo{author}{\bibfnamefont{P.~A.}
  \bibnamefont{Sundararajan}}, \bibinfo{journal}{Physical Review D (Particles,
  Fields, Gravitation, and Cosmology)} \textbf{\bibinfo{volume}{71}},
  \bibinfo{eid}{084008} (pages~\bibinfo{numpages}{16}) (\bibinfo{year}{2005}),
  \urlprefix\url{http://link.aps.org/abstract/PRD/v71/e084008}.

\bibitem[{\citenamefont{Buonanno
  et~al.}(2003{\natexlab{b}})\citenamefont{Buonanno, Chen, and
  Vallisneri}}]{BuonannoChenVallisneri:2003a}
\bibinfo{author}{\bibfnamefont{A.}~\bibnamefont{Buonanno}},
  \bibinfo{author}{\bibfnamefont{Y.}~\bibnamefont{Chen}}, \bibnamefont{and}
  \bibinfo{author}{\bibfnamefont{M.}~\bibnamefont{Vallisneri}},
  \bibinfo{journal}{Phys. Rev. D} \textbf{\bibinfo{volume}{67}},
  \bibinfo{pages}{024016} (\bibinfo{year}{2003}{\natexlab{b}}),
  \bibinfo{note}{erratum-ibid. 74 (2006) 029903(E)}.

\bibitem[{\citenamefont{Poisson and Will}(1995)}]{Poisson:1995ef}
\bibinfo{author}{\bibfnamefont{E.}~\bibnamefont{Poisson}} \bibnamefont{and}
  \bibinfo{author}{\bibfnamefont{C.~M.} \bibnamefont{Will}},
  \bibinfo{journal}{Phys. Rev.} \textbf{\bibinfo{volume}{D52}},
  \bibinfo{pages}{848} (\bibinfo{year}{1995}), \eprint{gr-qc/9502040}.

\bibitem[{\citenamefont{Lazzarini and Weiss}(1996)}]{LIGO-E950018-02-E}
\bibinfo{author}{\bibfnamefont{A.}~\bibnamefont{Lazzarini}} \bibnamefont{and}
  \bibinfo{author}{\bibfnamefont{R.}~\bibnamefont{Weiss}},
  \bibinfo{journal}{Technical document {LIGO}-E950018-02-E}
  (\bibinfo{year}{1996}).

\bibitem[{\citenamefont{{The {LIGO} Scientific
  Collaboration}}(2007)}]{LIGOS3S4Tuning}
\bibinfo{author}{\bibnamefont{{The {LIGO} Scientific Collaboration}}},
  \bibinfo{type}{Tech. Rep.} \bibinfo{number}{{LIGO}-T070109-01},
  \bibinfo{institution}{{LIGO} Project} (\bibinfo{year}{2007}),
  \urlprefix\url{http://www.ligo.caltech.edu/docs/T/T070109-01.pdf}.

\bibitem[{\citenamefont{Christensen et~al.}(2005)}]{Vetoes}
\bibinfo{author}{\bibfnamefont{N.}~\bibnamefont{Christensen}}
  \bibnamefont{et~al.} (\bibinfo{collaboration}{{LIGO} Scientific
  Collaboration}), \bibinfo{journal}{Class. Quant. Grav.}
  \textbf{\bibinfo{volume}{22}}, \bibinfo{pages}{S1059} (\bibinfo{year}{2005}).

\bibitem[{\citenamefont{Phinney}(1991)}]{Phinney:1991ei}
\bibinfo{author}{\bibfnamefont{E.~S.} \bibnamefont{Phinney}},
  \bibinfo{journal}{Astrophysical Journal} \textbf{\bibinfo{volume}{380}},
  \bibinfo{pages}{L17} (\bibinfo{year}{1991}).

\bibitem[{\citenamefont{Abbott et~al.}(2004{\natexlab{b}})}]{LIGOS1iul}
\bibinfo{author}{\bibfnamefont{B.}~\bibnamefont{Abbott}} \bibnamefont{et~al.}
  (\bibinfo{collaboration}{{LIGO} Scientific Collaboration}),
  \bibinfo{journal}{Phys. Rev. D} \textbf{\bibinfo{volume}{69}},
  \bibinfo{pages}{122001} (\bibinfo{year}{2004}{\natexlab{b}}),
  \eprint{gr-qc/0308069}.

\bibitem[{\citenamefont{Abbott et~al.}(2005)}]{LIGOS2iul}
\bibinfo{author}{\bibfnamefont{B.}~\bibnamefont{Abbott}} \bibnamefont{et~al.}
  (\bibinfo{collaboration}{{LIGO} Scientific Collaboration}),
  \bibinfo{journal}{Phys. Rev. D} \textbf{\bibinfo{volume}{72}},
  \bibinfo{pages}{082001} (\bibinfo{year}{2005}), \eprint{gr-qc/0505041}.

\bibitem[{\citenamefont{Brady et~al.}(2004)\citenamefont{Brady, Creighton, and
  Wiseman}}]{loudestGWDAW03}
\bibinfo{author}{\bibfnamefont{P.~R.} \bibnamefont{Brady}},
  \bibinfo{author}{\bibfnamefont{J.~D.~E.} \bibnamefont{Creighton}},
  \bibnamefont{and} \bibinfo{author}{\bibfnamefont{A.~G.}
  \bibnamefont{Wiseman}}, \bibinfo{journal}{Class. Quant. Grav.}
  \textbf{\bibinfo{volume}{21}}, \bibinfo{pages}{S1775} (\bibinfo{year}{2004}).

\bibitem[{\citenamefont{Biswas et~al.}(2007)\citenamefont{Biswas, Brady,
  Creighton, and Fairhurst}}]{ul}
\bibinfo{author}{\bibfnamefont{R.}~\bibnamefont{Biswas}},
  \bibinfo{author}{\bibfnamefont{P.~R.} \bibnamefont{Brady}},
  \bibinfo{author}{\bibfnamefont{J.~D.~E.} \bibnamefont{Creighton}},
  \bibnamefont{and} \bibinfo{author}{\bibfnamefont{S.}~\bibnamefont{Fairhurst}}
  (\bibinfo{year}{2007}), \eprint{arXiv:0710.0465 [gr-qc]}.

\bibitem[{\citenamefont{Brady and Fairhurst}(2007)}]{systematics}
\bibinfo{author}{\bibfnamefont{P.~R.} \bibnamefont{Brady}} \bibnamefont{and}
  \bibinfo{author}{\bibfnamefont{S.}~\bibnamefont{Fairhurst}}
  (\bibinfo{year}{2007}), \eprint{arXiv:0707.2410 [gr-qc]}.

\bibitem[{\citenamefont{{Kopparapu} et~al.}(2008)\citenamefont{{Kopparapu},
  {Hanna}, {Kalogera}, {O'Shaughnessy}, {Gonzalez}, {Brady}, and
  {Fairhurst}}}]{LIGOS3S4Galaxies}
\bibinfo{author}{\bibfnamefont{R.~K.} \bibnamefont{{Kopparapu}}},
  \bibinfo{author}{\bibfnamefont{C.~R.} \bibnamefont{{Hanna}}},
  \bibinfo{author}{\bibfnamefont{V.}~\bibnamefont{{Kalogera}}},
  \bibinfo{author}{\bibfnamefont{R.}~\bibnamefont{{O'Shaughnessy}}},
  \bibinfo{author}{\bibfnamefont{G.}~\bibnamefont{{Gonzalez}}},
  \bibinfo{author}{\bibfnamefont{P.~R.} \bibnamefont{{Brady}}},
  \bibnamefont{and}
  \bibinfo{author}{\bibfnamefont{S.}~\bibnamefont{{Fairhurst}}}
  (\bibinfo{year}{2008}), \eprint{due to appear in ApJ March 2008, 676}.

\bibitem[{\citenamefont{Press et~al.}(1992)\citenamefont{Press, Teukolsky,
  Vetterling, and Flannery}}]{NumericalRecipesInC}
\bibinfo{author}{\bibfnamefont{W.~H.} \bibnamefont{Press}},
  \bibinfo{author}{\bibfnamefont{S.~A.} \bibnamefont{Teukolsky}},
  \bibinfo{author}{\bibfnamefont{W.~T.} \bibnamefont{Vetterling}},
  \bibnamefont{and} \bibinfo{author}{\bibfnamefont{B.~P.}
  \bibnamefont{Flannery}}, \emph{\bibinfo{title}{Numerical Recipes in C: The
  Art of Scientiﬁc Computing}} (\bibinfo{publisher}{Cambridge University
  Press}, \bibinfo{address}{Cambridge, England}, \bibinfo{year}{1992}).

\end{thebibliography}

\end{document}